\documentclass[letter]{aa} 
\usepackage{graphicx}
\usepackage{txfonts}
\usepackage{natbib}     
\usepackage{xcolor} 
\usepackage{threeparttable} 
\usepackage{float}

\bibpunct{(}{)}{;}{a}{}{,} 

\begin{document}

   \title{Silicon in the dayside atmospheres of two ultra-hot Jupiters}

\author{D.~Cont\inst{1} 
        \and
        F.~Yan\inst{1}   
        \and
        A.~Reiners\inst{1}   
        \and
        L.~Nortmann\inst{1}  
        \and
    K.~Molaverdikhani\inst{2,3,4,5}  
        \and
        E.~Pall\'e\inst{6,7}  
        \and
        M.~Stangret\inst{6,7}   
        \and
        Th.~Henning\inst{5}    
        \and
        I.~Ribas\inst{8,9} 
    \and
        A.~Quirrenbach\inst{4}
        \and
        J.~A.~Caballero\inst{10}  
        \and
        M.~R.~Zapatero~Osorio\inst{11}   
        \and
        P.~J.~Amado\inst{12}  
        \and
    J.~Aceituno\inst{12,13}   
        \and  
        N.~Casasayas-Barris\inst{14} 
        \and
        S.~Czesla\inst{15,16} 
        \and
        A.~Kaminski\inst{4}
    \and
        M.~L\'opez-Puertas\inst{12}    
        \and
        D.~Montes\inst{17}  
        \and
    J.~C.~Morales\inst{8,9}  
        \and
        G.~Morello\inst{6,7}  
        \and
        E.~Nagel\inst{15,16} 
        \and 
        A.~S\'anchez-L\'opez\inst{14}
        \and
    E.~Sedaghati\inst{12,18}
        \and
        M.~Zechmeister\inst{1} 
\\
}

\institute{Institut f\"ur Astrophysik, Georg-August-Universit\"at, Friedrich-Hund-Platz 1, 37077 G\"ottingen, Germany\\
        \email{david.cont@uni-goettingen.de, fei.yan@uni-goettingen.de}
        \and
        Universit\"ats-Sternwarte, Ludwig-Maximilians-Universit\"at M\"unchen, Scheinerstrasse 1, 81679 M\"unchen, Germany
        \and
        Exzellenzcluster Origins, Boltzmannstraße 2, 85748 Garching, Germany
        \and
        Landessternwarte, Zentrum f\"ur Astronomie der Universit\"at Heidelberg, K\"onigstuhl 12, 69117 Heidelberg, Germany      
        \and
        Max-Planck-Institut f{\"u}r Astronomie, K{\"o}nigstuhl 17, 69117 Heidelberg, Germany     
        \and
        Instituto de Astrof{\'i}sica de Canarias (IAC), Calle V{\'i}a Lactea s/n, 38200 La Laguna, Tenerife, Spain
        \and
        Departamento de Astrof{\'i}sica, Universidad de La Laguna, 38026  La Laguna, Tenerife, Spain
        \and
        Institut de Ci\`encies de l'Espai (CSIC-IEEC), Campus UAB, c/ de Can Magrans s/n, 08193 Bellaterra, Barcelona, Spain
        \and
        Institut d'Estudis Espacials de Catalunya (IEEC), 08034 Barcelona, Spain
        \and
        Centro de Astrobiolog{\'i}a (CSIC-INTA), ESAC, Camino bajo del castillo s/n, 28692 Villanueva de la Ca{\~n}ada, Madrid, Spain
        \and
    Centro de Astrobiolog{\'i}a (CSIC-INTA), Carretera de Ajalvir km 4, E-28850 Torrej{\'o}n de Ardoz, Madrid, Spain
        \and
        Instituto de Astrof{\'i}sica de Andaluc{\'i}a (IAA-CSIC), Glorieta de la Astronom{\'i}a s/n, 18008 Granada, Spain
        \and
    Centro Astron{\'o}nomico Hispano Alem{\'a}n, Observatorio de Calar Alto, Sierra de los Filabres, E-04550 G{\'e}rgal, Spain
    \and
        Leiden Observatory, Universiteit Leiden, Postbus 9513, 2300 RA, Leiden, The Netherlands
        \and
        Hamburger Sternwarte, Universit{\"a}t Hamburg, Gojenbergsweg 112, 21029 Hamburg, Germany
        \and
        Th{\"u}ringer Landessternwarte Tautenburg, Sternwarte 5, 07778 Tautenburg, Germany
        \and
        Departamento de F\'{i}sica de la Tierra y Astrof\'{i}sica 
        and IPARCOS-UCM (Instituto de F\'{i}sica de Part\'{i}culas y del Cosmos de la UCM), 
        Facultad de Ciencias F\'{i}sicas, Universidad Complutense de Madrid, E-28040, Madrid, Spain
        \and
        Facultad de Ingenier\'ia y Ciencias, Universidad Adolfo Ib\'a\~nez, Av.\ Diagonal las Torres 2640, Pe\~nalol\'en, Santiago, Chile
        \\      
}

\date{Received 29 November 2021 / Accepted 9 December 2021}


\abstract
{Atmospheres of highly irradiated gas giant planets host a large variety of atomic and ionic species. Here we observe the thermal emission spectra of the two ultra-hot Jupiters WASP-33b and KELT-20b/MASCARA-2b in the near-infrared wavelength range with CARMENES. Via high-resolution Doppler spectroscopy, we searched for neutral silicon (Si) in their dayside atmospheres. We detect the Si spectral signature of both planets via cross-correlation with model spectra. Detection levels of 4.8$\mathrm{\sigma}$ and 5.4$\mathrm{\sigma}$, respectively, are observed when assuming a solar atmospheric composition. This is the first detection of Si in exoplanet atmospheres. The presence of Si is an important finding due to its fundamental role in cloud formation and, hence, for the planetary energy balance. Since the spectral lines are detected in emission, our results also confirm the presence of an inverted temperature profile in the dayside atmospheres of both planets.}

\maketitle

%

\section{Introduction}

        Ultra-hot Jupiters (UHJs) are highly irradiated gas giant planets with equilibrium temperatures ($T_\mathrm{eq}$) close to the stellar regime ($T_\mathrm{eq}$\,$\ge$\,2200\,K; \citealt{Parmentier2018}). Planets in this regime are expected to be tidally locked to their host stars, given enough time for tidal forces to synchronize the rotation of the planet to its orbital motion. The extreme thermal conditions in combination with permanent day- and nightsides allow the existence of a large variety of chemical species. In the dayside atmospheres of UHJs, most of the molecules are expected to be dissociated, leading to the presence of atomic and ionic species \citep[e.g.,][]{Lothringer2018, Arcangeli2018, Kitzmann2018}. Molecules should be widely present in atmospheric regions other than the dayside, spanning from bi-atomic molecules at the terminators to more complex compounds on the planetary nightsides \citep{Helling2019}. To date, various chemical species have been detected in the transmission or emission spectra of UHJs. This includes atomic hydrogen and metals such as Ca, Cr, Fe, Mg, Mn, Na, Ti, Sc, V, and Y \citep[e.g.,][]{Fossati2010, Jensen2018, Yan&Henning2018, Casasayas-Barris2018, Casasayas-Barris2019, Hoeijmakers2018, Hoeijmakers2019, Hoeijmakers2020, Sing2019, Cauley2019, Cauley2021, Stangret2020, Nugroho2020_Fe,Nugroho2020_KELT20b, BenYami2020, Borsa2021_2, Tabernero2021, Yan_2021_submitted, Yan2021} and molecules such as H$_2$O and OH\footnote{For a more complete list of detections, we refer the reader to the \texttt{Exoplanet Atmospheres Database} available at \texttt{http://research.iac.es/proyecto/exoatmospheres/}.} \citep[e.g.,][]{Huitson2013, Edwards2020, Tsiaras2018, Mikal-Evans2020, Nugroho2021}.
        
        Inverted temperature-pressure ($T$-$p$) profiles have been measured in a number of UHJ atmospheres \citep[e.g.,][]{Haynes2015, Evans2017, Sheppard2017, Arcangeli2018, Kreidberg2018, Mansfield2018, Nugroho2020_Fe, Yan2020}. The presence of these so-called temperature inversions (i.e., temperature increasing with altitude) was initially explained via strong absorption of the incoming stellar radiation by TiO and VO \citep{Hubeny2003, Fortney2008}. However, the presence of TiO is under debate due to the conflicting results of different studies \citep{Evans2016, Nugroho2017, Herman2020, Edwards2020, Serindag2021, Cont2021}. Besides, the search for VO remains elusive at high spectral resolution \citep{Merritt2020}. On the other hand, as various atoms and ions have been discovered in UHJs with an inverted atmosphere, atomic species have become promising candidates for causing and maintaining temperature inversions \citep{Lothringer2018, Hoeijmakers2020_WASP-121b}.
        
        Theoretical simulations predict the presence of Si in the atmospheres of UHJs. Atomic Si is expected in planetary daysides, while Si-bearing molecules (e.g., SiO) should be prominent on the nightsides and in the terminator regions \citep{Helling2019}. Si is suggested to play a key role in cloud formation, which strongly impacts the atmospheric energy budget \citep{Gao2020, Gao&Poweell2021}. The abundance of Si is expected to be close to that of Fe for a solar atmospheric composition \citep{Fossati2021}.
        
        Hints for atmospheric \ion{Si}{iii} absorption in the 1206.5\,$\AA$ resonance line were found via transmission spectroscopy in HD~209458b \citep{Linsky2010}. However, \cite{Ballester2015} later disproved this detection by identifying stellar flux variations as the cause for a false positive signal in the data. In addition to this specific spectral line, \cite{Fossati2021} proposed investigating the spectral features of \ion{Si}{ii} around 1530\,$\AA$ as a further way of searching for this atomic species in planetary atmospheres. \cite{Hoeijmakers2019} searched for Si in KELT-9b via transmission spectroscopy at high spectral resolution but did not detect it. This non-detection is probably due to a low concentration of neutral Si in the atmosphere of KELT-9b that is, in turn, due to the strong ionization of the species. Ionized Si should be present, but its spectral signature is expected to be featureless in the investigated wavelength range.
        
        In this Letter we report the first detection of neutral Si in exoplanet atmospheres. We detected \ion{Si}{i} emission lines in the dayside atmospheres of WASP-33b and KELT-20b/MASCARA-2b via high-resolution emission spectroscopy in the near-infrared.
        WASP-33b \citep{Collier-Cameron2010} orbits an A-type star with $\delta$ Scuti pulsations \citep{Herrero2011}. With $T_\mathrm{eq}$\,$\sim$\,2700\,K, it is the second hottest planet known to date, showing a temperature inversion in its dayside atmosphere \citep{Haynes2015}. The spectral features of \ion{Ca}{ii}, \ion{Fe}{i}, OH, TiO, and the hydrogen Balmer lines have been found at high spectral resolution \citep{Nugroho2017, Yan2019, Nugroho2020_Fe, Cauley2021, Yan2021, Borsa2021_1, Nugroho2021, Cont2021}, and AlO and FeH were tentatively detected \citep{Essen2019, Kesseli2020}. KELT-20b/MASCARA-2b \citep{Lund2017, Talens2018} is a UHJ with $T_\mathrm{eq}$\,$\sim$\,2300\,K that orbits an A-type star without pulsations. A number of metals, such as Ca, Cr, Fe, Na, and Mg, were found in the planetary transmission spectrum \citep{Casasayas-Barris2018, Casasayas-Barris2019, Hoeijmakers2020, Stangret2020, Nugroho2020_KELT20b, Rainer2021}. Recently, \cite{Yan_2021_submitted} used the spectral emission lines of \ion{Fe}{i} to retrieve the atmospheric temperature profile, claiming the presence of a temperature inversion on the planetary dayside. The parameters of both planetary systems are summarized in Table~\ref{tab-parameters}.

%

\section{Observations and data reduction}
\label{Observations and data reduction}

%
\begin{table*}
        \caption{Observation log.}             
        \label{obs_log}      
        \centering                          
        \begin{threeparttable}
                \begin{tabular}{l l l l l l l }        
                        \hline\hline                 
                        \noalign{\smallskip}
                        Object & Date & Observing time & Airmass change & Phase coverage & Exposure time & $N_\mathrm{spectra}$  \\     
                        \noalign{\smallskip}
                        \hline                       
                        \noalign{\smallskip}
                        WASP-33b & 2017-11-15 & 18:13--04:04\,UT         & 1.74--1.00--1.95 &  0.29--0.63  & 300\,s & 88\tablefootmark{a}\\  
                        KELT-20b & 2020-05-21 & 23:05--03:07\,UT         & 1.87--1.02       &  0.41--0.46  & 125\,s & 85\\  
                        KELT-20b & 2020-07-09 & 23:00--03:05\,UT         & 1.07--1.01--1.17 &  0.51--0.56  & 125\,s & 85\\                       
                        \noalign{\smallskip}
                        \hline                                   
                \end{tabular}
                \tablefoot{
                        \tablefoottext{a}{Total number of spectra is 105; 17 spectra with insufficient quality were removed.}                     
                }
        \end{threeparttable}  
        
\end{table*}

\begin{figure}
        \centering
        \includegraphics[width=0.5\textwidth]{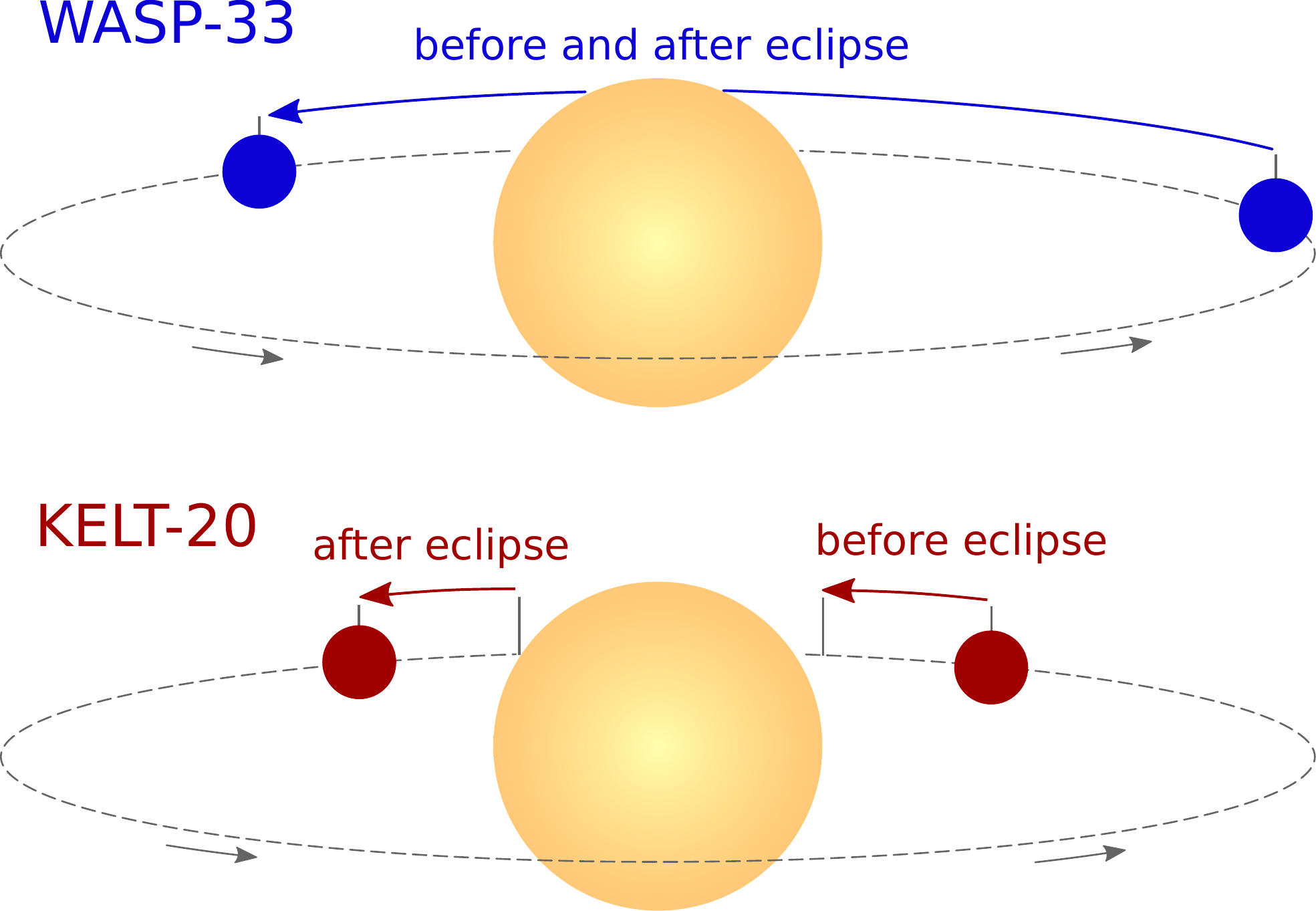}
        \caption{Schematic of emission spectroscopy observations. The orbital phase coverage of WASP-33b is shown in the {\it top panel} (night of 15 November 2017) and that of KELT-20b in the {\it bottom panel} (before 21 May 2020 and after 9 July 2020). The orbital motion direction is indicated by the arrows.}
        \label{orbital_phase_coverage}
\end{figure}

We observed the two planets over a total of three nights with the CARMENES (Calar Alto high-Resolution search for M dwarfs with Exoearths with Near-infrared and optical \'Echelle Spectrographs) spectrograph at the Calar Alto Observatory \citep{Quirrenbach2014, Quirrenbach2020}. The observation of WASP-33b took place on 15~November~2017, and KELT-20b was observed on two nights, 21~May~2020 and 9~July~2020 (see Fig.~\ref{orbital_phase_coverage}). To investigate the dayside atmospheres, we observed at orbital phases close to the secondary eclipse. These observations have already been utilized in previous studies \citep{Cont2021, Yan_2021_submitted} to retrieve the Fe emission spectrum in the visible channel (5200--9600\,$\AA$; $R$\,$\sim$\,94,600). In this work we analyze the data collected with the near-infrared channel (9600--17,100\,$\AA$; $R$\,$\sim$\,80,400), for which the Si signal is expected to be stronger (see the model spectra with solar Si abundance in Sect.~\ref{Model spectra}). For seven WASP-33b spectra, the star was not centered on the fiber, and for three spectra the flux dropped due to a passing cloud. For another seven spectra, the elevation of the target was too low (airmass\,>\,2) to get a useful amount of flux in the near-infrared channel. Hence, we removed a total of 17 spectra from the WASP-33b observations. The targets were observed with fiber~A, and fiber~B was used to record the sky background. Details of the observations are given in the observation log in Table~\ref{obs_log}.

We used the reduction pipeline {\tt caracal}~v2.20 \citep{Zechmeister2014, Caballero2016} to extract the order-by-order\footnote{In the near-infrared channel, two detectors are located along the dispersion direction. The resulting spectra are therefore split into half-orders. For simplicity, we use the term "spectral order" to describe these "half-orders" in the following text.} one-dimensional spectra and the corresponding uncertainties from the raw frames. The data of each night were reduced separately. We excluded the spectra from the echelle orders 45-43 due to an insufficient flux level. These orders correspond to the water absorption band around 1.4\,$\mu$m, for which the Earth's atmosphere is almost entirely opaque. For each spectral order, we arranged the spectra chronologically to obtain the two-dimensional spectral matrix (Fig.~\ref{SYSREM}a). To account for the variable continuum level, we fitted the individual spectra with a second-order polynomial and normalized them with the fit function. Wavelength ranges with strong emission lines in fiber~B were excluded from the second-order polynomial fit. The outliers were removed by applying 5$\sigma$ clipping to the time evolution of each pixel. We masked the wavelength bins with flux below 20\,\% of the continuum level (Fig.~\ref{SYSREM}b).

To remove the telluric and stellar lines from the spectra, we used the detrending algorithm \texttt{SYSREM} \citep{Tamuz2005}. We passed the normalized spectral matrix and the corresponding uncertainties as an input to the algorithm (details in Appendix~\ref{Removal of telluric and stellar lines with SYSREM}). The uncertainties were computed via error propagation. We ran \texttt{SYSREM} for ten consecutive times, resulting in a residual spectral matrix for each iteration. If present in the data, the signature of the planetary atmosphere is buried in the noise of the residual spectra (Fig.~\ref{SYSREM}c).

%

\section{Method}
\label{Detection of neutral Si}

\subsection{Model spectra}
\label{Model spectra}

\begin{figure*}
        \centering
        \includegraphics[width=\textwidth]{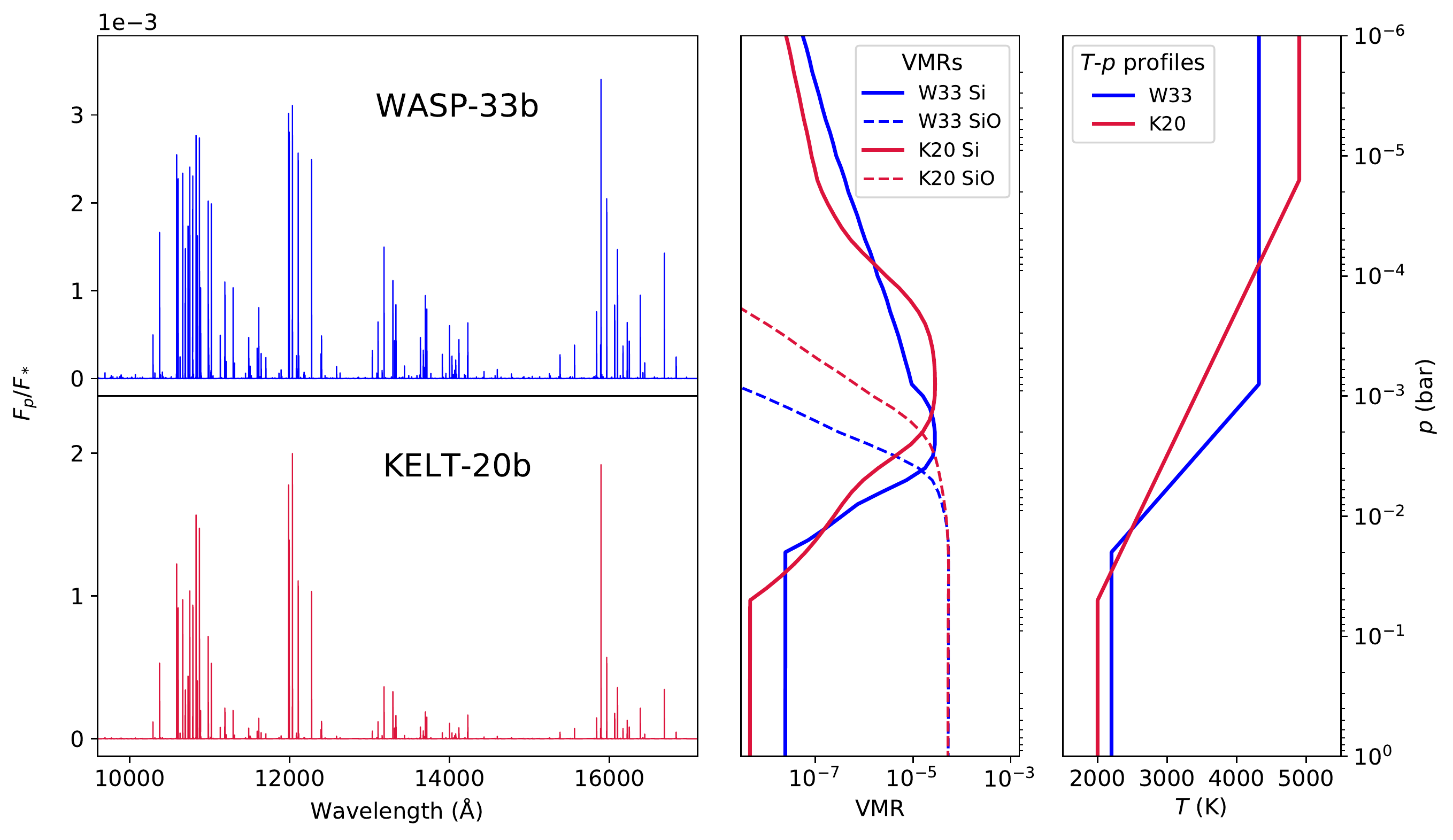}
        \caption{Emission model spectra ({\it left panels}) for WASP-33b (W33, blue) and KELT-20b (K20, red) and their corresponding VMRs of Si and SiO ({\it middle panel}) and $T$-$p$ profiles ({\it right panel}).
        We assumed equilibrium chemistry and [Si/H]\,=\,0 to generate the presented model spectra (for sub- and super-solar metallicity values, see Figs.~\ref{model_spectra_W33_appendix} and \ref{model_spectra_K20_appendix}).} 
        \label{model_spectra}
\end{figure*}

The model atmosphere of each planet was divided into 61 layers, evenly spaced on a logarithmic pressure scale from 1 to $10^{-6}$\,bar. For WASP-33b, we adopted the $T$-$p$ profile of WASP-189b from \cite{Yan2020}, which was retrieved via the \ion{Fe}{i} emission spectrum and by assuming a solar metallicity. This choice is motivated by the physical similarities between the two planets. The profile was also successfully used in a prior work to detect the \ion{Fe}{i} signature in the atmosphere of WASP-33b  \citep{Cont2021}. For KELT-20b, we took the $T$-$p$ profile from a joint retrieval of CARMENES and TESS (Transiting Exoplanet Survey Satellite) by \cite{Yan_2021_submitted}. We deployed \texttt{easyCHEM} \citep{Molliere2017} to compute the volume mixing ratio (VMR) and the mean molecular weight of each atmospheric layer. To this end, we assumed equilibrium chemistry, at five different values of metallicity [M/H] between \mbox{--2\,dex} and +2\,dex in steps of 1\,dex. We assumed that all metals vary with overall metallicity and, hence, [Si/H]\,=\,[M/H]. Figure~\ref{model_spectra} shows that under the assumption of equilibrium chemistry, neutral Si is most abundant at the location of the thermal inversion layers. Deeper in the atmospheres, SiO accounts for the majority of Si inventory. At higher altitudes, the VMR of the species decreases due to ionization.

We used the radiative transfer code \texttt{petitRADTRANS} \citep{Molliere2019} to generate the model spectra. The continuum opacity of H$^-$ was not taken into account, as it was found to only insignificantly affect the resulting model spectra (see Fig.~\ref{model_spectra_compare_H}). The Si opacities for the radiative transfer calculation were computed from the Kurucz line database \citep{Kurucz2018}. For each planet, this resulted in five model emission spectra with different Si abundances (see Figs.~\ref{model_spectra_W33_appendix} and \ref{model_spectra_K20_appendix}).

We computed the planet-to-star flux ratio of each model spectrum by dividing by the blackbody spectrum of the respective host star. As the reduced spectra were normalized, we also normalized the model spectra to the continuum. After convolving with the instrumental profile, we obtained the final emission model spectra for cross-correlation. The model spectra with [Si/H]\,=\,0 are shown in Fig.~\ref{model_spectra}.

\subsection{Cross-correlation}
\label{Cross-correlation}

\begin{figure*}
        \centering
        \includegraphics[width=\textwidth]{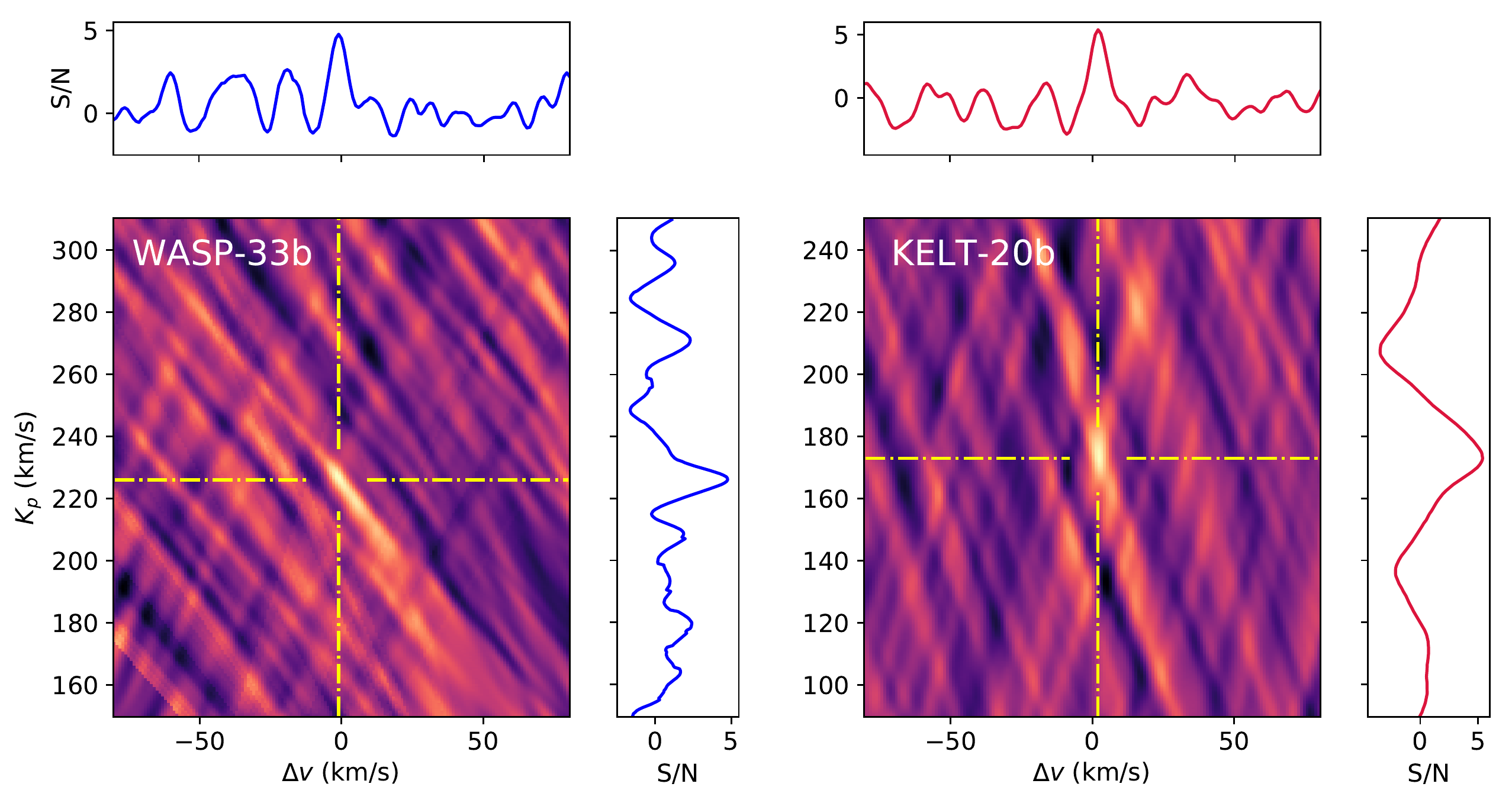}
        \caption{S/N detection maps of neutral Si for WASP-33b ({\it left panel}) and KELT-20b ({\it right panel}). The signal of WASP-33b peaks with a S/N of 4.8 after three consecutive \texttt{SYSREM} iterations. For KELT-20b, we achieve the highest significance at S/N\,=\,5.4 after four iterations. We indicate the peak coordinates by the yellow dashed-dotted lines. The horizontal and vertical panels correspond to the cross sections of the S/N peaks.}
        \label{SN_maps}
\end{figure*}

The model spectra in Sect.~\ref{Model spectra} predict a planet-to-star flux ratio on the order of $10^{-3}$ or lower. Hence, the planetary signal is dominated by noise in the \texttt{SYSREM} reduced spectra. We applied the cross-correlation method to extract the emission signature of the planetary atmosphere \citep[e.g.,][]{Snellen2010, Brogi2012, Alonso-Floriano2019, Sanchez-Lopez2019}. This technique maps the planetary emission lines onto a single peak, enabling the identification of the species in the planetary signal. We performed the cross-correlation analysis for each of the model spectra with different Si abundance separately.

The model spectrum was shifted over a radial velocity (RV) range from --520\,km\,s$^{-1}$ to +520\,km\,s$^{-1}$ with steps of 1\,km\,s$^{-1}$. At each step, we multiplied the shifted model spectrum with the uncertainty-weighted residual spectra. As a result, we obtained the weighted cross-correlation function (CCF), defined as
\begin{equation}
      \mathrm{CCF} = \sum r_i m_i(\varv),
\end{equation}
for each observed spectrum and echelle order \citep{Gibson2020}. We denote with $r_i$ the residual spectra weighted by the inverse of the squared uncertainties; $m_i$ is the model spectrum shifted by $\varv$ in the RV space. For each echelle order the CCFs were stacked into an individual array. Subsequently, we co-added the arrays from different echelle orders, leading to the final CCF map for each spectral model and observation night. Finally, we merged the CCF maps of the two KELT-20b observations.

The stellar line profile of WASP-33 undergoes time-dependent variations due to the pulsations of the star \citep{Herrero2011}. Lines of neutral Si are also present in the stellar spectrum and, consequently, not efficiently removed by \texttt{SYSREM}. This causes the pulsations of the star to appear as artifacts in the CCF map \citep{Nugroho2020_Fe, Cont2021}. To exclude potential spurious signals from the pulsations of WASP-33, we masked the RV range between $\pm \varv_\mathrm{rot}\sin i_*$ (i.e., between --87\,km\,s$^{-1}$ and +87\,km\,s$^{-1}$) in the stellar rest frame \citep{Cont2021}. In contrast, KELT-20 has no pulsations, which results in an efficient stellar line removal by \texttt{SYSREM}. No masking was therefore required in the CCF map of KELT-20b.

For each planet, we aligned the CCF map to the planetary rest frame over a range of different orbital velocity semi-amplitudes ($K_\mathrm{p}$). We assumed a circular orbit with a planetary RV of
\begin{equation}
\label{equ-orb-v}
\varv_\mathrm{p} = \varv_\mathrm{sys} + \varv_\mathrm{bary} + K_\mathrm{p} \sin\left(2\pi\phi\right) + \Delta \varv
\end{equation}
for shifting the CCF map, with $\varv_\mathrm{sys}$ the systemic velocity, $\varv_\mathrm{bary}$ the barycentric velocity of the observer, $\Delta \varv$ the velocity deviation from the planetary rest frame, and $\phi$ the orbital phase. For each value of $K_\mathrm{p}$, we collapsed the CCF map into a one-dimensional CCF by calculating the mean value over all orbital phases. The CCFs from different $K_\mathrm{p}$ values were stacked in a two-dimensional array, which was further normalized by its standard deviation (excluding the region around the strongest signal peak). This resulted in a signal-to-noise mapping of the detection significance (S/N map), which enabled us to assess the presence of Si in the planetary atmospheres.

\section{Results and discussion}
\label{Results and discussion}

We detected the spectral signature of neutral Si in the dayside atmospheres of the two exoplanets WASP-33b and KELT-20b. At \texttt{SYSREM} iterations higher than one and for all tested metallicity values ([Si/H] between --2\,dex and +2\,dex in steps of 1\,dex; see Sect.~\ref{Model spectra}), the signal is identified in the S/N maps. For each planet, the strongest signal was found assuming an atmosphere with a solar Si abundance. The respective S/N maps are shown in Fig.~\ref{SN_maps}. For non-solar Si abundances, the detection peaks are less prominent. We show the S/N maps for non-solar abundances in Figs.~\ref{SN_different_SiH_W33_appendix} and \ref{SN_different_SiH_K20_appendix}. The evolution of the S/N with increasing \texttt{SYSREM} iterations is plotted in Fig.~\ref{SN_iterations}, and Fig.~\ref{CCF_trails} shows the aligned planetary trails together with the profile of the detection peaks. In Appendix \ref{Null detection test}, we also show that our implementation of the cross-correlation technique does not lead to significant detection peaks when an inappropriate model spectrum is used.

For WASP-33b, we found the most significant detection after three \texttt{SYSREM} iterations at S/N\,=\,4.8. The peak is located at $K_\mathrm{p}$\,=\,$226.0_{-11.5}^{+5.0}$\,km\,s$^{-1}$, which is close to the expected $K_\mathrm{p}$ of $231\pm3$\,km\,s$^{-1}$ calculated from the orbital parameters of the planet \citep{Kovacs2013, Lehmann2015}. For $\Delta \varv$ we find a small value of $-1.0_{-4.0}^{+10.0}$\,km\,s$^{-1}$, which is consistent with zero. The $K_\mathrm{p}$ of our detection is slightly lower than the expected value, a trend that has also been found in prior studies of Fe \citep{Nugroho2020_Fe, Cont2021}. We also detected the spectral signature of neutral Si in the atmosphere of KELT-20b. The strongest signal was found after four consecutive \texttt{SYSREM} iterations with a S/N of 5.4 at $K_\mathrm{p}$\,=\,$173.0_{-5.0}^{+6.5}$\,km\,s$^{-1}$ and $\Delta \varv$\,=\,$2.0_{-2.0}^{+2.0}$\,km\,s$^{-1}$. This result agrees with the $K_\mathrm{p}$ values of $173.4_{-1.5}^{+1.8}$\,km\,s$^{-1}$ and $169.3_{-4.6}^{+5.9}$\,km\,s$^{-1}$ calculated from the system parameters of \cite{Talens2018} and \cite{Lund2017}, respectively. The small value of $\Delta \varv$ is also consistent with zero. All results are summarized in Table~\ref{tab-results}.

We also investigated whether the spectral lines of Si are affected by rotational broadening. For each planet, we simulated two CCFs: the auto-correlation of the non-broadened model spectrum and the cross-correlation between the non-broadened model and a rotationally broadened model. We assumed a tidally locked rotation, corresponding to rotation velocities of 7\,km\,s$^{-1}$ and 3\,km\,s$^{-1}$ at the equators of WASP-33b and KELT-20b, respectively. Figure~\ref{CCF_trails} compares the profile of the measured CCF peaks with the simulated detection peaks. For WASP-33b, the observed CCF is best reproduced when no rotation is assumed, which hints toward a localized distribution of Si in the planetary atmosphere. However, for KELT-20b, the difference between the broadened and non-broadened simulations is marginal, and both of them are consistent with the observed CCF, which indicates that the rotational broadening probably makes a negligible contribution to the total line profile.

We detected the spectral lines of the species in emission, which confirms the presence of thermal inversion layers in the dayside atmospheres of WASP-33b \citep{Nugroho2017, Nugroho2020_Fe, Cont2021} and KELT-20b \citep{Yan_2021_submitted}. Together with recent detections of neutral Fe \citep[e.g.,][]{Pino2020, Yan2020, Nugroho2020_Fe, Cont2021, Kasper2021}, the presence of neutral Si also strengthens the assumption that atomic species play a key role in the energy balance of UHJ atmospheres. Due to a comparable ionization potential, we expect similar VMRs of Si and Fe in the upper atmosphere of UHJs \citep{Fossati2021}. However, the detection of Si is more challenging than that of Fe because of the smaller number of significant emission lines.

Although our detections are strongest when assuming a solar Si abundance, there is a degeneracy between the metallicity and the selected $T$-$p$ profiles. Hence, our result of a solar metallicity in both planets is only valid for the specific $T$-$p$ profiles that were selected. Considering the model spectra with [Si/H]\,=\,0 in Figs.~\ref{model_spectra_W33_appendix} and \ref{model_spectra_K20_appendix}, we also conclude that the most prominent Si features in the planetary spectrum are probably restricted to the wavelength interval 10,000--13,000\,$\AA$ and to a small region around 16,000\,$\AA$.

\begin{table}
        \caption{Summary of results.} 
        \label{tab-results} 
        \centering     
    \renewcommand{\arraystretch}{1.2} 
        \begin{threeparttable}
                \begin{tabular}{l l l l}      
                        \hline\hline  
                        \noalign{\smallskip}
                         Object & S/N  & $K_\mathrm{p}$ (km\,s$^{-1}$) & $\Delta \varv$ (km\,s$^{-1}$)   \\           
                        \noalign{\smallskip}
                        \hline    

                         WASP-33b     & 4.8    & $226.0_{-11.5}^{+5.0}$  & $-1.0_{-4.0}^{+10.0}$  \\     
                         KELT-20b     & 5.4    & $173.0_{-5.0}^{+6.5}$   & $2.0_{-2.0}^{+2.0}$    \\     
                        \noalign{\smallskip}
                        \hline                               
                \end{tabular}
        \end{threeparttable}      
\end{table}

We note that in the S/N maps, detection peaks can even be observed at the lowest metallicity values investigated. This is due to the fact that the cross-correlation technique only takes the strength of the spectral lines relative to one another into account, not their absolute strength. The information about the absolute value of the CCFs is removed by the normalization step that is included in the calculation of the S/N maps (see Sect.~\ref{Cross-correlation}). As shown in Figs.~\ref{model_spectra_W33_appendix} and \ref{model_spectra_K20_appendix}, the model spectra at [Si/H]\,$\le$\,0 have spectral lines with a similar strength relative to one another. Therefore, it is plausible that the model spectra of sub-solar metallicities cause a similar S/N detection pattern compared to a solar metallicity despite their weak emission lines.

Neutral Si was not detected in the HARPS-N transmission spectra of KELT-9b \citep{Hoeijmakers2019}. This is not surprising, since Si is probably largely ionized due to the extreme thermal conditions in the atmosphere of this planet. In fact, \cite{Fossati2021} predicted that Si begins to get ionized at pressures around $10^{-2}$\,bar in the atmosphere of KELT-9b. In addition, the transmission spectrum of ionized Si is expected to be featureless in the investigated wavelength range. Consequently, for planetary atmospheres with extreme thermal conditions such as KELT-9b, Si may be difficult to detect. For planets with more moderate thermal conditions, we suggest that the search for Si could be limited due to Si depletion. In this scenario, most of the Si would be bound in SiO and other Si-bearing molecules, which can also condense out of the gas phase. We therefore posit that the thermal conditions in exoplanet atmospheres may be a crucial constraint for the search of Si.

Si is supposed to be an important element for cloud formation in exoplanet atmospheres, with silicates dominating the cloud composition over a wide range of planetary equilibrium temperatures. Our detections of Si in its gaseous phase indicate that the dayside atmospheres of UHJs are hotter than the condensation temperatures of Si-bearing condensates. This is in line with theoretical work that predicts the presence of silicate clouds primarily on the planetary nightsides \citep{Gao2020, Gao&Poweell2021}.

%

\section{Conclusions}
\label{Conclusions}

We used the CARMENES spectrograph to observe the dayside emission spectra of two UHJs -- WASP-33b and KELT-20b. By using the cross-correlation technique, we detected the signature of neutral Si 
in the exoplanet atmospheres. For both planets, the Doppler shifts of their Si spectra are consistent with the known orbital motion. We tested model spectra with different Si abundances and detected the strongest signals when assuming a solar abundance for the planetary atmospheres. From our Si model spectra, we conclude that the presence of prominent spectral features is probably restricted to two narrow regions in the near-infrared wavelength range. The spectral lines of Si were detected in emission, which is unambiguous evidence for the existence of temperature inversions in the two planetary atmospheres.

In combination with the presence of Fe, reported in prior studies, our detections of Si suggest that atomic species play a key role in the atmospheric heating process that is necessary to maintain a thermal inversion layer. Strong absorption lines of ionized Si should exist in the ultraviolet transmission spectra of UHJs and may be detectable with the Hubble Space Telescope and the upcoming World Space Observatory-Ultraviolet. Future observations over a wider wavelength range will provide further constraints on Si in planetary atmospheres, with the potential of shedding light on complex processes of cloud formation.

%

\begin{acknowledgements}

CARMENES is an instrument at the Centro Astron\'omico Hispano-Alem\'an (CAHA) at Calar Alto (Almer\'{\i}a, Spain), operated jointly by the Junta de Andaluc\'ia and the Instituto de Astrof\'isica de Andaluc\'ia (CSIC).
        
        CARMENES was funded by the Max-Planck-Gesellschaft (MPG), 
        the Consejo Superior de Investigaciones Cient\'{\i}ficas (CSIC),
        the Ministerio de Econom\'ia y Competitividad (MINECO) and the European Regional Development Fund (ERDF) through projects FICTS-2011-02, ICTS-2017-07-CAHA-4, and CAHA16-CE-3978, 
        and the members of the CARMENES Consortium 
        (Max-Planck-Institut f\"ur Astronomie,
        Instituto de Astrof\'{\i}sica de Andaluc\'{\i}a,
        Landessternwarte K\"onigstuhl,
        Institut de Ci\`encies de l'Espai,
        Institut f\"ur Astrophysik G\"ottingen,
        Universidad Complutense de Madrid,
        Th\"uringer Landessternwarte Tautenburg,
        Instituto de Astrof\'{\i}sica de Canarias,
        Hamburger Sternwarte,
        Centro de Astrobiolog\'{\i}a and
        Centro Astron\'omico Hispano-Alem\'an), 
        with additional contributions by the MINECO, 
        the Deutsche Forschungsgemeinschaft through the Major Research Instrumentation Programme and Research Unit FOR2544 ``Blue Planets around Red Stars'', 
        the Klaus Tschira Stiftung, 
        the states of Baden-W\"urttemberg and Niedersachsen, 
        and by the Junta de Andaluc\'{\i}a.
        
        We acknowledge financial support from the
        Deutsche Forschungsgemeinschaft through the priority program SPP 1992 ``Exploring the Diversity of Extrasolar Planets'' (RE 1664/16-1), and the Research Unit FOR2544 ``Blue Planets around Red Stars'' (RE 1664/21-1). T.H. acknowledges support from the European Research Council under the Horizon 2020 Framework Program via the ERC Advanced Grant Origins 83 24 28. G.M. has received funding from the European Union's Horizon 2020 research and innovation programme under the Marie Sk\l{}odowska-Curie grant agreement No. 895525. E.S. acknowledges support from ANID - Millennium Science Initiative - ICN12\_009. A.S.L. acknowledges funding from the European Research Council under the European Union's Horizon 2020 research and innovation program under grant agreement No. 694513. This research was supported by the Excellence Cluster ORIGINS which is funded by the Deutsche Forschungsgemeinschaft (DFG, German Research Foundation) under Germany's Excellence Strategy - EXC-2094 - 390783311.

\end{acknowledgements}

\bibliographystyle{aa} 

\bibliography{Si-K20-W33-refer}

\begin{thebibliography}{78}
\expandafter\ifx\csname natexlab\endcsname\relax\def\natexlab#1{#1}\fi

\bibitem[{{Alonso-Floriano} {et~al.}(2019){Alonso-Floriano},
  {S{\'a}nchez-L{\'o}pez}, {Snellen}, {L{\'o}pez-Puertas}, {Nagel}, {Amado},
  {Bauer}, {Caballero}, {Czesla}, {Nortmann}, {Pall{\'e}}, {Salz}, {Reiners},
  {Ribas}, {Quirrenbach}, {Aceituno}, {Anglada-Escud{\'e}}, {B{\'e}jar},
  {Guenther}, {Henning}, {Kaminski}, {K{\"u}rster}, {Lamp{\'o}n}, {Lara},
  {Montes}, {Morales}, {Tal-Or}, {Schmitt}, {Zapatero Osorio}, \&
  {Zechmeister}}]{Alonso-Floriano2019}
{Alonso-Floriano}, F.~J., {S{\'a}nchez-L{\'o}pez}, A., {Snellen}, I.~A.~G.,
  {et~al.} 2019, \aap, 621, A74

\bibitem[{{Arcangeli} {et~al.}(2018){Arcangeli}, {D{\'e}sert}, {Line}, {Bean},
  {Parmentier}, {Stevenson}, {Kreidberg}, {Fortney}, {Mansfield}, \&
  {Showman}}]{Arcangeli2018}
{Arcangeli}, J., {D{\'e}sert}, J.-M., {Line}, M.~R., {et~al.} 2018, \apj, 855,
  L30

\bibitem[{{Ballester} \& {Ben-Jaffel}(2015)}]{Ballester2015}
{Ballester}, G.~E. \& {Ben-Jaffel}, L. 2015, \apj, 804, 116

\bibitem[{{Ben-Yami} {et~al.}(2020){Ben-Yami}, {Madhusudhan}, {Cabot},
  {Constantinou}, {Piette}, {Gandhi}, \& {Welbanks}}]{BenYami2020}
{Ben-Yami}, M., {Madhusudhan}, N., {Cabot}, S. H.~C., {et~al.} 2020, \apjl,
  897, L5

\bibitem[{{Borsa} {et~al.}(2021{\natexlab{a}}){Borsa}, {Allart},
  {Casasayas-Barris}, {Tabernero}, {Zapatero Osorio}, {Cristiani}, {Pepe},
  {Rebolo}, {Santos}, {Adibekyan}, {Bourrier}, {Demangeon}, {Ehrenreich},
  {Pall{\'e}}, {Sousa}, {Lillo-Box}, {Lovis}, {Micela}, {Oshagh}, {Poretti},
  {Sozzetti}, {Allende Prieto}, {Alibert}, {Amate}, {Benz}, {Bouchy}, {Cabral},
  {Dekker}, {D'Odorico}, {Di Marcantonio}, {Figueira}, {Genova Santos},
  {Gonz{\'a}lez Hern{\'a}ndez}, {Lo Curto}, {Manescau}, {Martins},
  {M{\'e}gevand}, {Mehner}, {Molaro}, {Nunes}, {Riva}, {Su{\'a}rez
  Mascare{\~n}o}, {Udry}, \& {Zerbi}}]{Borsa2021_2}
{Borsa}, F., {Allart}, R., {Casasayas-Barris}, N., {et~al.} 2021{\natexlab{a}},
  \aap, 645, A24

\bibitem[{{Borsa} {et~al.}(2021{\natexlab{b}}){Borsa}, {Lanza}, {Raspantini},
  {Rainer}, {Fossati}, {Brogi}, {Di Mauro}, {Gratton}, {Pino}, {Benatti},
  {Bignamini}, {Bonomo}, {Claudi}, {Esposito}, {Frustagli}, {Maggio},
  {Maldonado}, {Mancini}, {Micela}, {Nascimbeni}, {Poretti}, {Scandariato},
  {Sicilia}, {Sozzetti}, {Boschin}, {Cosentino}, {Covino}, {Desidera}, {Di
  Fabrizio}, {Fiorenzano}, {Harutyunyan}, {Knapic}, {Molinari}, {Pagano},
  {Pedani}, \& {Piotto}}]{Borsa2021_1}
{Borsa}, F., {Lanza}, A.~F., {Raspantini}, I., {et~al.} 2021{\natexlab{b}},
  \aap, 653, A104

\bibitem[{{Brogi} {et~al.}(2012){Brogi}, {Snellen}, {de Kok}, {Albrecht},
  {Birkby}, \& {de Mooij}}]{Brogi2012}
{Brogi}, M., {Snellen}, I. A.~G., {de Kok}, R.~J., {et~al.} 2012, \nat, 486,
  502

\bibitem[{{Caballero} {et~al.}(2016){Caballero}, {Gu{\`a}rdia}, {L{\'o}pez del
  Fresno}, {Zechmeister}, {de Juan}, {Alonso-Floriano}, {Amado}, {Colom{\'e}},
  {Cort{\'e}s-Contreras}, {Garc{\'{\i}}a-Piquer}, {Gesa}, {de Guindos},
  {Hagen}, {Helmling}, {Hern{\'a}ndez Casta{\~n}o}, {K{\"u}rster},
  {L{\'o}pez-Santiago}, {Montes}, {Morales Mu{\~n}oz}, {Pavlov}, {Quirrenbach},
  {Reiners}, {Ribas}, {Seifert}, \& {Solano}}]{Caballero2016}
{Caballero}, J.~A., {Gu{\`a}rdia}, J., {L{\'o}pez del Fresno}, M., {et~al.}
  2016, in \procspie, Vol. 9910, Observatory Operations: Strategies, Processes,
  and Systems VI, 99100E

\bibitem[{{Casasayas-Barris} {et~al.}(2018){Casasayas-Barris}, {Pall{\'e}},
  {Yan}, {Chen}, {Albrecht}, {Nortmann}, {Van Eylen}, {Snellen}, {Talens},
  {Gonz{\'a}lez Hern{\'a}ndez}, {Rebolo}, \& {Otten}}]{Casasayas-Barris2018}
{Casasayas-Barris}, N., {Pall{\'e}}, E., {Yan}, F., {et~al.} 2018, \aap, 616,
  A151

\bibitem[{{Casasayas-Barris} {et~al.}(2019){Casasayas-Barris}, {Pall{\'e}},
  {Yan}, {Chen}, {Kohl}, {Stangret}, {Parviainen}, {Helling}, {Watanabe},
  {Czesla}, {Fukui}, {Monta{\~n}{\'e}s-Rodr{\'\i}guez}, {Nagel}, {Narita},
  {Nortmann}, {Nowak}, {Schmitt}, \& {Zapatero Osorio}}]{Casasayas-Barris2019}
{Casasayas-Barris}, N., {Pall{\'e}}, E., {Yan}, F., {et~al.} 2019, \aap, 628,
  A9

\bibitem[{{Cauley} {et~al.}(2019){Cauley}, {Shkolnik}, {Ilyin}, {Strassmeier},
  {Redfield}, \& {Jensen}}]{Cauley2019}
{Cauley}, P.~W., {Shkolnik}, E.~L., {Ilyin}, I., {et~al.} 2019, \aj, 157, 69

\bibitem[{{Cauley} {et~al.}(2021){Cauley}, {Wang}, {Shkolnik}, {Ilyin},
  {Strassmeier}, {Redfield}, \& {Jensen}}]{Cauley2021}
{Cauley}, P.~W., {Wang}, J., {Shkolnik}, E.~L., {et~al.} 2021, \aj, 161, 152

\bibitem[{{Collier Cameron} {et~al.}(2010){Collier Cameron}, {Guenther},
  {Smalley}, {McDonald}, {Hebb}, {Andersen}, {Augusteijn}, {Barros}, {Brown},
  {Cochran}, {Endl}, {Fossey}, {Hartmann}, {Maxted}, {Pollacco}, {Skillen},
  {Telting}, {Waldmann}, \& {West}}]{Collier-Cameron2010}
{Collier Cameron}, A., {Guenther}, E., {Smalley}, B., {et~al.} 2010, \mnras,
  407, 507

\bibitem[{{Cont} {et~al.}(2021){Cont}, {Yan}, {Reiners}, {Casasayas-Barris},
  {Molli{\`e}re}, {Pall{\'e}}, {Henning}, {Nortmann}, {Stangret}, {Czesla},
  {L{\'o}pez-Puertas}, {S{\'a}nchez-L{\'o}pez}, {Rodler}, {Ribas},
  {Quirrenbach}, {Caballero}, {Amado}, {Carone}, {Khaimova}, {Kreidberg},
  {Molaverdikhani}, {Montes}, {Morello}, {Nagel}, {Oshagh}, \&
  {Zechmeister}}]{Cont2021}
{Cont}, D., {Yan}, F., {Reiners}, A., {et~al.} 2021, \aap, 651, A33

\bibitem[{{Edwards} {et~al.}(2020){Edwards}, {Changeat}, {Baeyens}, {Tsiaras},
  {Al-Refaie}, {Taylor}, {Yip}, {Bieger}, {Blain}, {Gressier}, {Guilluy},
  {Jaziri}, {Kiefer}, {Modirrousta-Galian}, {Morvan}, {Mugnai}, {Pluriel},
  {Poveda}, {Skaf}, {Whiteford}, {Wright}, {Zingales}, {Charnay}, {Drossart},
  {Leconte}, {Venot}, {Waldmann}, \& {Beaulieu}}]{Edwards2020}
{Edwards}, B., {Changeat}, Q., {Baeyens}, R., {et~al.} 2020, \aj, 160, 8

\bibitem[{{Evans} {et~al.}(2017){Evans}, {Sing}, {Kataria}, {Goyal}, {Nikolov},
  {Wakeford}, {Deming}, {Marley}, {Amundsen}, {Ballester}, {Barstow},
  {Ben-Jaffel}, {Bourrier}, {Buchhave}, {Cohen}, {Ehrenreich}, {Garc{\'\i}a
  Mu{\~n}oz}, {Henry}, {Knutson}, {Lavvas}, {Lecavelier Des Etangs}, {Lewis},
  {L{\'o}pez-Morales}, {Mandell}, {Sanz-Forcada}, {Tremblin}, \&
  {Lupu}}]{Evans2017}
{Evans}, T.~M., {Sing}, D.~K., {Kataria}, T., {et~al.} 2017, \nat, 548, 58

\bibitem[{{Evans} {et~al.}(2016){Evans}, {Sing}, {Wakeford}, {Nikolov},
  {Ballester}, {Drummond}, {Kataria}, {Gibson}, {Amundsen}, \&
  {Spake}}]{Evans2016}
{Evans}, T.~M., {Sing}, D.~K., {Wakeford}, H.~R., {et~al.} 2016, \apjl, 822, L4

\bibitem[{{Fortney} {et~al.}(2008){Fortney}, {Lodders}, {Marley}, \&
  {Freedman}}]{Fortney2008}
{Fortney}, J.~J., {Lodders}, K., {Marley}, M.~S., \& {Freedman}, R.~S. 2008,
  \apj, 678, 1419

\bibitem[{{Fossati} {et~al.}(2010){Fossati}, {Haswell}, {Froning}, {Hebb},
  {Holmes}, {Kolb}, {Helling}, {Carter}, {Wheatley}, {Collier Cameron},
  {Loeillet}, {Pollacco}, {Street}, {Stempels}, {Simpson}, {Udry}, {Joshi},
  {West}, {Skillen}, \& {Wilson}}]{Fossati2010}
{Fossati}, L., {Haswell}, C.~A., {Froning}, C.~S., {et~al.} 2010, \apj, 714,
  L222

\bibitem[{{Fossati} {et~al.}(2021){Fossati}, {Young}, {Shulyak}, {Koskinen},
  {Huang}, {Cubillos}, {France}, \& {Sreejith}}]{Fossati2021}
{Fossati}, L., {Young}, M.~E., {Shulyak}, D., {et~al.} 2021, \aap, 653, A52

\bibitem[{{Gao} \& {Powell}(2021)}]{Gao&Poweell2021}
{Gao}, P. \& {Powell}, D. 2021, \apjl, 918, L7

\bibitem[{{Gao} {et~al.}(2020){Gao}, {Thorngren}, {Lee}, {Fortney}, {Morley},
  {Wakeford}, {Powell}, {Stevenson}, \& {Zhang}}]{Gao2020}
{Gao}, P., {Thorngren}, D.~P., {Lee}, E. K.~H., {et~al.} 2020, Nature
  Astronomy, 4, 951

\bibitem[{{Gibson} {et~al.}(2020){Gibson}, {Merritt}, {Nugroho}, {Cubillos},
  {de Mooij}, {Mikal-Evans}, {Fossati}, {Lothringer}, {Nikolov}, {Sing},
  {Spake}, {Watson}, \& {Wilson}}]{Gibson2020}
{Gibson}, N.~P., {Merritt}, S., {Nugroho}, S.~K., {et~al.} 2020, \mnras, 493,
  2215

\bibitem[{{Haynes} {et~al.}(2015){Haynes}, {Mandell}, {Madhusudhan}, {Deming},
  \& {Knutson}}]{Haynes2015}
{Haynes}, K., {Mandell}, A.~M., {Madhusudhan}, N., {Deming}, D., \& {Knutson},
  H. 2015, \apj, 806, 146

\bibitem[{{Helling} {et~al.}(2019){Helling}, {Gourbin}, {Woitke}, \&
  {Parmentier}}]{Helling2019}
{Helling}, C., {Gourbin}, P., {Woitke}, P., \& {Parmentier}, V. 2019, \aap,
  626, A133

\bibitem[{{Herman} {et~al.}(2020){Herman}, {de Mooij}, {Jayawardhana}, \&
  {Brogi}}]{Herman2020}
{Herman}, M.~K., {de Mooij}, E. J.~W., {Jayawardhana}, R., \& {Brogi}, M. 2020,
  \aj, 160, 93

\bibitem[{{Herrero} {et~al.}(2011){Herrero}, {Morales}, {Ribas}, \&
  {Naves}}]{Herrero2011}
{Herrero}, E., {Morales}, J.~C., {Ribas}, I., \& {Naves}, R. 2011, \aap, 526,
  L10

\bibitem[{{Hoeijmakers} {et~al.}(2020{\natexlab{a}}){Hoeijmakers}, {Cabot},
  {Zhao}, {Buchhave}, {Tronsgaard}, {Davis}, {Kitzmann}, {Grimm}, {Cegla},
  {Bourrier}, {Ehrenreich}, {Heng}, {Lovis}, \& {Fischer}}]{Hoeijmakers2020}
{Hoeijmakers}, H.~J., {Cabot}, S. H.~C., {Zhao}, L., {et~al.}
  2020{\natexlab{a}}, \aap, 641, A120

\bibitem[{{Hoeijmakers} {et~al.}(2018){Hoeijmakers}, {Ehrenreich}, {Heng},
  {Kitzmann}, {Grimm}, {Allart}, {Deitrick}, {Wyttenbach}, {Oreshenko}, {Pino},
  {Rimmer}, {Molinari}, \& {Di Fabrizio}}]{Hoeijmakers2018}
{Hoeijmakers}, H.~J., {Ehrenreich}, D., {Heng}, K., {et~al.} 2018, \nat, 560,
  453

\bibitem[{{Hoeijmakers} {et~al.}(2019){Hoeijmakers}, {Ehrenreich}, {Kitzmann},
  {Allart}, {Grimm}, {Seidel}, {Wyttenbach}, {Pino}, {Nielsen}, {Fisher},
  {Rimmer}, {Bourrier}, {Cegla}, {Lavie}, {Lovis}, {Patzer}, {Stock}, {Pepe},
  \& {Heng}}]{Hoeijmakers2019}
{Hoeijmakers}, H.~J., {Ehrenreich}, D., {Kitzmann}, D., {et~al.} 2019, \aap,
  627, A165

\bibitem[{{Hoeijmakers} {et~al.}(2020{\natexlab{b}}){Hoeijmakers}, {Seidel},
  {Pino}, {Kitzmann}, {Sindel}, {Ehrenreich}, {Oza}, {Bourrier}, {Allart},
  {Gebek}, {Lovis}, {Yurchenko}, {Astudillo-Defru}, {Bayliss}, {Cegla},
  {Lavie}, {Lendl}, {Melo}, {Murgas}, {Nascimbeni}, {Pepe}, {S{\'e}gransan},
  {Udry}, {Wyttenbach}, \& {Heng}}]{Hoeijmakers2020_WASP-121b}
{Hoeijmakers}, H.~J., {Seidel}, J.~V., {Pino}, L., {et~al.} 2020{\natexlab{b}},
  \aap, 641, A123

\bibitem[{{Hubeny} {et~al.}(2003){Hubeny}, {Burrows}, \&
  {Sudarsky}}]{Hubeny2003}
{Hubeny}, I., {Burrows}, A., \& {Sudarsky}, D. 2003, \apj, 594, 1011

\bibitem[{{Huitson} {et~al.}(2013){Huitson}, {Sing}, {Pont}, {Fortney},
  {Burrows}, {Wilson}, {Ballester}, {Nikolov}, {Gibson}, {Deming}, {Aigrain},
  {Evans}, {Henry}, {Lecavelier des Etangs}, {Showman}, {Vidal-Madjar}, \&
  {Zahnle}}]{Huitson2013}
{Huitson}, C.~M., {Sing}, D.~K., {Pont}, F., {et~al.} 2013, \mnras, 434, 3252

\bibitem[{{Jensen} {et~al.}(2018){Jensen}, {Cauley}, {Redfield}, {Cochran}, \&
  {Endl}}]{Jensen2018}
{Jensen}, A.~G., {Cauley}, P.~W., {Redfield}, S., {Cochran}, W.~D., \& {Endl},
  M. 2018, \aj, 156, 154

\bibitem[{{Johnson} {et~al.}(2015){Johnson}, {Cochran}, {Collier Cameron}, \&
  {Bayliss}}]{Johnson2015}
{Johnson}, M.~C., {Cochran}, W.~D., {Collier Cameron}, A., \& {Bayliss}, D.
  2015, \apj, 810, L23

\bibitem[{{Kasper} {et~al.}(2021){Kasper}, {Bean}, {Line}, {Seifahrt},
  {St{\"u}rmer}, {Pino}, {D{\'e}sert}, \& {Brogi}}]{Kasper2021}
{Kasper}, D., {Bean}, J.~L., {Line}, M.~R., {et~al.} 2021, \apjl, 921, L18

\bibitem[{{Kesseli} {et~al.}(2020){Kesseli}, {Snellen}, {Alonso-Floriano},
  {Molli{\`e}re}, \& {Serindag}}]{Kesseli2020}
{Kesseli}, A.~Y., {Snellen}, I.~A.~G., {Alonso-Floriano}, F.~J.,
  {Molli{\`e}re}, P., \& {Serindag}, D.~B. 2020, \aj, 160, 228

\bibitem[{{Kitzmann} {et~al.}(2018){Kitzmann}, {Heng}, {Rimmer}, {Hoeijmakers},
  {Tsai}, {Malik}, {Lendl}, {Deitrick}, \& {Demory}}]{Kitzmann2018}
{Kitzmann}, D., {Heng}, K., {Rimmer}, P.~B., {et~al.} 2018, \apj, 863, 183

\bibitem[{{Kov{\'a}cs} {et~al.}(2013){Kov{\'a}cs}, {Kov{\'a}cs}, {Hartman},
  {Bakos}, {Bieryla}, {Latham}, {Noyes}, {Reg{\'a}ly}, \&
  {Esquerdo}}]{Kovacs2013}
{Kov{\'a}cs}, G., {Kov{\'a}cs}, T., {Hartman}, J.~D., {et~al.} 2013, \aap, 553,
  A44

\bibitem[{{Kreidberg} {et~al.}(2018){Kreidberg}, {Line}, {Parmentier},
  {Stevenson}, {Louden}, {Bonnefoy}, {Faherty}, {Henry}, {Williamson},
  {Stassun}, {Beatty}, {Bean}, {Fortney}, {Showman}, {D{\'e}sert}, \&
  {Arcangeli}}]{Kreidberg2018}
{Kreidberg}, L., {Line}, M.~R., {Parmentier}, V., {et~al.} 2018, \aj, 156, 17

\bibitem[{{Kurucz}(2018)}]{Kurucz2018}
{Kurucz}, R.~L. 2018, in Astronomical Society of the Pacific Conference Series,
  Vol. 515, Workshop on Astrophysical Opacities, 47

\bibitem[{{Lehmann} {et~al.}(2015){Lehmann}, {Guenther}, {Sebastian},
  {D{\"o}llinger}, {Hartmann}, \& {Mkrtichian}}]{Lehmann2015}
{Lehmann}, H., {Guenther}, E., {Sebastian}, D., {et~al.} 2015, \aap, 578, L4

\bibitem[{{Linsky} {et~al.}(2010){Linsky}, {Yang}, {France}, {Froning},
  {Green}, {Stocke}, \& {Osterman}}]{Linsky2010}
{Linsky}, J.~L., {Yang}, H., {France}, K., {et~al.} 2010, \apj, 717, 1291

\bibitem[{{Lothringer} {et~al.}(2018){Lothringer}, {Barman}, \&
  {Koskinen}}]{Lothringer2018}
{Lothringer}, J.~D., {Barman}, T., \& {Koskinen}, T. 2018, \apj, 866, 27

\bibitem[{{Lund} {et~al.}(2017){Lund}, {Rodriguez}, {Zhou}, {Gaudi}, {Stassun},
  {Johnson}, {Bieryla}, {Oelkers}, {Stevens}, {Collins}, {Penev}, {Quinn},
  {Latham}, {Villanueva}, {Eastman}, {Kielkopf}, {Oberst}, {Jensen}, {Cohen},
  {Joner}, {Stephens}, {Relles}, {Corfini}, {Gregorio}, {Zambelli}, {Esquerdo},
  {Calkins}, {Berlind}, {Ciardi}, {Dressing}, {Patel}, {Gagnon}, {Gonzales},
  {Beatty}, {Siverd}, {Labadie-Bartz}, {Kuhn}, {Col{\'o}n}, {James}, {Pepper},
  {Fulton}, {McLeod}, {Stockdale}, {Calchi Novati}, {DePoy}, {Gould},
  {Marshall}, {Trueblood}, {Trueblood}, {Johnson}, {Wright}, {McCrady},
  {Wittenmyer}, {Johnson}, {Sergi}, {Wilson}, \& {Sliski}}]{Lund2017}
{Lund}, M.~B., {Rodriguez}, J.~E., {Zhou}, G., {et~al.} 2017, \aj, 154, 194

\bibitem[{{Maciejewski} {et~al.}(2018){Maciejewski}, {Fern{\'a}ndez},
  {Aceituno}, {Mart{\'\i}n-Ruiz}, {Ohlert}, {Dimitrov}, {Szyszka}, {von Essen},
  {Mugrauer}, {Bischoff}, {Michel}, {Mallonn}, {Stangret}, \&
  {Mo{\'z}dzierski}}]{Maciejewski2018}
{Maciejewski}, G., {Fern{\'a}ndez}, M., {Aceituno}, F., {et~al.} 2018, \actaa,
  68, 371

\bibitem[{{Mansfield} {et~al.}(2018){Mansfield}, {Bean}, {Line}, {Parmentier},
  {Kreidberg}, {D{\'e}sert}, {Fortney}, {Stevenson}, {Arcangeli}, \&
  {Dragomir}}]{Mansfield2018}
{Mansfield}, M., {Bean}, J.~L., {Line}, M.~R., {et~al.} 2018, \aj, 156, 10

\bibitem[{{Merritt} {et~al.}(2020){Merritt}, {Gibson}, {Nugroho}, {de Mooij},
  {Hooton}, {Matthews}, {McKemmish}, {Mikal-Evans}, {Nikolov}, {Sing}, {Spake},
  \& {Watson}}]{Merritt2020}
{Merritt}, S.~R., {Gibson}, N.~P., {Nugroho}, S.~K., {et~al.} 2020, \aap, 636,
  A117

\bibitem[{{Mikal-Evans} {et~al.}(2020){Mikal-Evans}, {Sing}, {Kataria},
  {Wakeford}, {Mayne}, {Lewis}, {Barstow}, \& {Spake}}]{Mikal-Evans2020}
{Mikal-Evans}, T., {Sing}, D.~K., {Kataria}, T., {et~al.} 2020, \mnras, 496,
  1638

\bibitem[{{Molli{\`e}re} {et~al.}(2017){Molli{\`e}re}, {van Boekel}, {Bouwman},
  {Henning}, {Lagage}, \& {Min}}]{Molliere2017}
{Molli{\`e}re}, P., {van Boekel}, R., {Bouwman}, J., {et~al.} 2017, \aap, 600,
  A10

\bibitem[{{Molli{\`e}re} {et~al.}(2019){Molli{\`e}re}, {Wardenier}, {van
  Boekel}, {Henning}, {Molaverdikhani}, \& {Snellen}}]{Molliere2019}
{Molli{\`e}re}, P., {Wardenier}, J.~P., {van Boekel}, R., {et~al.} 2019, \aap,
  627, A67

\bibitem[{{Nugroho} {et~al.}(2020{\natexlab{a}}){Nugroho}, {Gibson}, {de
  Mooij}, {Herman}, {Watson}, {Kawahara}, \& {Merritt}}]{Nugroho2020_Fe}
{Nugroho}, S.~K., {Gibson}, N.~P., {de Mooij}, E. J.~W., {et~al.}
  2020{\natexlab{a}}, \apjl, 898, L31

\bibitem[{{Nugroho} {et~al.}(2020{\natexlab{b}}){Nugroho}, {Gibson}, {de
  Mooij}, {Watson}, {Kawahara}, \& {Merritt}}]{Nugroho2020_KELT20b}
{Nugroho}, S.~K., {Gibson}, N.~P., {de Mooij}, E. J.~W., {et~al.}
  2020{\natexlab{b}}, \mnras, 496, 504

\bibitem[{{Nugroho} {et~al.}(2021){Nugroho}, {Kawahara}, {Gibson}, {de Mooij},
  {Hirano}, {Kotani}, {Kawashima}, {Masuda}, {Brogi}, {Birkby}, {Watson},
  {Tamura}, {Zwintz}, {Harakawa}, {Kudo}, {Kuzuhara}, {Hodapp}, {Ishizuka},
  {Jacobson}, {Konishi}, {Kurokawa}, {Nishikawa}, {Omiya}, {Serizawa}, {Ueda},
  \& {Vievard}}]{Nugroho2021}
{Nugroho}, S.~K., {Kawahara}, H., {Gibson}, N.~P., {et~al.} 2021, \apjl, 910,
  L9

\bibitem[{{Nugroho} {et~al.}(2017){Nugroho}, {Kawahara}, {Masuda}, {Hirano},
  {Kotani}, \& {Tajitsu}}]{Nugroho2017}
{Nugroho}, S.~K., {Kawahara}, H., {Masuda}, K., {et~al.} 2017, \aj, 154, 221

\bibitem[{{Parmentier} {et~al.}(2018){Parmentier}, {Line}, {Bean}, {Mansfield},
  {Kreidberg}, {Lupu}, {Visscher}, {D{\'e}sert}, {Fortney}, {Deleuil},
  {Arcangeli}, {Showman}, \& {Marley}}]{Parmentier2018}
{Parmentier}, V., {Line}, M.~R., {Bean}, J.~L., {et~al.} 2018, \aap, 617, A110

\bibitem[{{Pino} {et~al.}(2020){Pino}, {D{\'e}sert}, {Brogi}, {Malavolta},
  {Wyttenbach}, {Line}, {Hoeijmakers}, {Fossati}, {Bonomo}, {Nascimbeni},
  {Panwar}, {Affer}, {Benatti}, {Biazzo}, {Bignamini}, {Borsa}, {Carleo},
  {Claudi}, {Cosentino}, {Covino}, {Damasso}, {Desidera}, {Giacobbe},
  {Harutyunyan}, {Lanza}, {Leto}, {Maggio}, {Maldonado}, {Mancini}, {Micela},
  {Molinari}, {Pagano}, {Piotto}, {Poretti}, {Rainer}, {Scandariato},
  {Sozzetti}, {Allart}, {Borsato}, {Bruno}, {Di Fabrizio}, {Ehrenreich},
  {Fiorenzano}, {Frustagli}, {Lavie}, {Lovis}, {Magazz{\`u}}, {Nardiello},
  {Pedani}, \& {Smareglia}}]{Pino2020}
{Pino}, L., {D{\'e}sert}, J.-M., {Brogi}, M., {et~al.} 2020, \apjl, 894, L27

\bibitem[{{Quirrenbach} {et~al.}(2014){Quirrenbach}, {Amado}, {Caballero},
  {Mundt}, {Reiners}, {Ribas}, {Seifert}, {Abril}, {Aceituno},
  {Alonso-Floriano}, {Ammler-von Eiff}, {Antona Jim{\'e}nez},
  {Anwand-Heerwart}, {Azzaro}, {Bauer}, {Barrado}, {Becerril}, {B{\'e}jar},
  {Ben{\'\i}tez}, {Berdi{\~n}as}, {C{\'a}rdenas}, {Casal}, {Claret},
  {Colom{\'e}}, {Cort{\'e}s-Contreras}, {Czesla}, {Doellinger}, {Dreizler},
  {Feiz}, {Fern{\'a}ndez}, {Galad{\'\i}}, {G{\'a}lvez-Ortiz},
  {Garc{\'\i}a-Piquer}, {Garc{\'\i}a-Vargas}, {Garrido}, {Gesa}, {G{\'o}mez
  Galera}, {Gonz{\'a}lez {\'A}lvarez}, {Gonz{\'a}lez Hern{\'a}ndez},
  {Gr{\"o}zinger}, {Gu{\`a}rdia}, {Guenther}, {de Guindos},
  {Guti{\'e}rrez-Soto}, {Hagen}, {Hatzes}, {Hauschildt}, {Helmling}, {Henning},
  {Hermann}, {Hern{\'a}ndez Casta{\~n}o}, {Herrero}, {Hidalgo}, {Holgado},
  {Huber}, {Huber}, {Jeffers}, {Joergens}, {de Juan}, {Kehr}, {Klein},
  {K{\"u}rster}, {Lamert}, {Lalitha}, {Laun}, {Lemke}, {Lenzen}, {L{\'o}pez del
  Fresno}, {L{\'o}pez Mart{\'\i}}, {L{\'o}pez-Santiago}, {Mall}, {Mandel},
  {Mart{\'\i}n}, {Mart{\'\i}n-Ruiz}, {Mart{\'\i}nez-Rodr{\'\i}guez}, {Marvin},
  {Mathar}, {Mirabet}, {Montes}, {Morales Mu{\~n}oz}, {Moya}, {Naranjo},
  {Ofir}, {Oreiro}, {Pall{\'e}}, {Panduro}, {Passegger}, {P{\'e}rez-Calpena},
  {P{\'e}rez Medialdea}, {Perger}, {Pluto}, {Ram{\'o}n}, {Rebolo}, {Redondo},
  {Reffert}, {Reinhardt}, {Rhode}, {Rix}, {Rodler}, {Rodr{\'\i}guez},
  {Rodr{\'\i}guez-L{\'o}pez}, {Rodr{\'\i}guez-P{\'e}rez}, {Rohloff}, {Rosich},
  {S{\'a}nchez-Blanco}, {S{\'a}nchez Carrasco}, {Sanz-Forcada}, {Sarmiento},
  {Sch{\"a}fer}, {Schiller}, {Schmidt}, {Schmitt}, {Solano}, {Stahl}, {Storz},
  {St{\"u}rmer}, {Su{\'a}rez}, {Ulbrich}, {Veredas}, {Wagner}, {Winkler},
  {Zapatero Osorio}, {Zechmeister}, {Abell{\'a}n de Paco},
  {Anglada-Escud{\'e}}, {del Burgo}, {Klutsch}, {Lizon}, {L{\'o}pez-Morales},
  {Morales}, {Perryman}, {Tulloch}, \& {Xu}}]{Quirrenbach2014}
{Quirrenbach}, A., {Amado}, P.~J., {Caballero}, J.~A., {et~al.} 2014, in
  Society of Photo-Optical Instrumentation Engineers (SPIE) Conference Series,
  Vol. 9147, Ground-based and Airborne Instrumentation for Astronomy V, ed.
  S.~K. {Ramsay}, I.~S. {McLean}, \& H.~{Takami}, 91471F

\bibitem[{{Quirrenbach} {et~al.}(2020){Quirrenbach}, {CARMENES Consortium},
  {Amado}, {Ribas}, {Reiners}, {Caballero}, {Aceituno}, {Alacid},
  {Alonso-Floriano}, {Anglada-Escud{\'e}}, {Azzaro}, {Baroch}, {Bauer},
  {Becerril}, {B{\'e}jar}, {Bluhm}, {Calvo Ortega}, {Cardona Guill{\'e}n},
  {Casasayas-Barris}, {Chaturvedi}, {Cifuentes}, {Colom{\'e}}, {Conte},
  {Cort{\'e}s-Contreras}, {Czesla}, {D{\'\i}ez-Alonso}, {Dom{\'\i}nguez
  Fern{\'a}ndez}, {Dreizler}, {Duque-Arribas}, {Espinoza}, {Fuhrmeister},
  {Galad{\'\i}-Enr{\'\i}quez}, {Gar{\'c}a Quintana}, {Gonz{\'a}lez-Alvare},
  {Gonz{\'a}lez Cuesta}, {Gonz{\'a}lez Hern{\'a}ndez}, {Guenther}, {de
  Guindos}, {Hatzes}, {Henning}, {Herbort}, {Herrero}, {Hintz},
  {Iglesias-P{\'a}ra}, {Jeffers}, {Johnson}, {de Juan}, {Kaminski}, {Kemmer},
  {Khaimova}, {Khalafinejad}, {Klahr}, {Kossakowski}, {Kreidberg},
  {K{\"u}rster}, {Labarga}, {Lafarga}, {Lamp{\'o}n}, {Lara}, {Lillo-Box},
  {Lodieu}, {L{\'o}pez Gallifa}, {L{\'o}pez Gonz{\'a}lez}, {L{\'o}pez-Puertas},
  {Luque}, {Marfil}, {Mart{\'\i}n-Ruiz}, {Matth{\'e}}, {Molaverdikhani},
  {Montes}, {Morales}, {Morales-Calder{\'o}on}, {Nagel}, {Nortmann}, {Nowak},
  {Ofir}, {Oshaghi}, {Pall{\'e}}, {Passegger}, {Pavlov}, {Pedraz},
  {Perdelwitz}, {Perger}, {Reffert}, {Revilla}, {Rodr{\'\i}guez},
  {Rodr{\'\i}guez L{\'o}pez}, {Sabotta}, {Sadegi}, {Sairam}, {Salz},
  {S{\'a}nchez-L{\'o}pez}, {Sanz-Forcada}, {Sarkis}, {Sch{\"a}fer}, {Schiller},
  {Schlecker}, {Schmitt}, {Sch{\"o}fer}, {Schweitzer}, {Seiferta}, {Shan},
  {Shulyak}, {Skrzypinski}, {Solano}, {Soto}, {Stahl}, {Stangret}, {Stock},
  {Strachan}, {Stuber}, {St{\"u}rmer}, {Tabernero}, {Tal-Or}, {Tala-Pinto},
  {Trifonov}, {Vanaverbeke}, {Yan}, {Zapatero Osorio}, \&
  {Zechmeister}}]{Quirrenbach2020}
{Quirrenbach}, A., {CARMENES Consortium}, {Amado}, P.~J., {et~al.} 2020, in
  Society of Photo-Optical Instrumentation Engineers (SPIE) Conference Series,
  Vol. 11447, Society of Photo-Optical Instrumentation Engineers (SPIE)
  Conference Series, 114473C

\bibitem[{{Rainer} {et~al.}(2021){Rainer}, {Borsa}, {Pino}, {Frustagli},
  {Brogi}, {Biazzo}, {Bonomo}, {Carleo}, {Claudi}, {Gratton}, {Lanza},
  {Maggio}, {Maldonado}, {Mancini}, {Micela}, {Scandariato}, {Sozzetti},
  {Buchschacher}, {Cosentino}, {Covino}, {Ghedina}, {Gonzalez}, {Leto}, {Lodi},
  {Martinez Fiorenzano}, {Molinari}, {Molinaro}, {Nardiello}, {Oliva},
  {Pagano}, {Pedani}, {Piotto}, \& {Poretti}}]{Rainer2021}
{Rainer}, M., {Borsa}, F., {Pino}, L., {et~al.} 2021, \aap, 649, A29

\bibitem[{{S{\'a}nchez-L{\'o}pez} {et~al.}(2019){S{\'a}nchez-L{\'o}pez},
  {Alonso-Floriano}, {L{\'o}pez-Puertas}, {Snellen}, {Funke}, {Nagel}, {Bauer},
  {Amado}, {Caballero}, {Czesla}, {Nortmann}, {Pall{\'e}}, {Salz}, {Reiners},
  {Ribas}, {Quirrenbach}, {Anglada-Escud{\'e}}, {B{\'e}jar},
  {Casasayas-Barris}, {Galad{\'\i}-Enr{\'\i}quez}, {Guenther}, {Henning},
  {Kaminski}, {K{\"u}rster}, {Lamp{\'o}n}, {Lara}, {Montes}, {Morales},
  {Stangret}, {Tal-Or}, {Sanz-Forcada}, {Schmitt}, {Zapatero Osorio}, \&
  {Zechmeister}}]{Sanchez-Lopez2019}
{S{\'a}nchez-L{\'o}pez}, A., {Alonso-Floriano}, F.~J., {L{\'o}pez-Puertas}, M.,
  {et~al.} 2019, \aap, 630, A53

\bibitem[{{Serindag} {et~al.}(2021){Serindag}, {Nugroho}, {Molli{\`e}re}, {de
  Mooij}, {Gibson}, \& {Snellen}}]{Serindag2021}
{Serindag}, D.~B., {Nugroho}, S.~K., {Molli{\`e}re}, P., {et~al.} 2021, \aap,
  645, A90

\bibitem[{{Sheppard} {et~al.}(2017){Sheppard}, {Mandell}, {Tamburo}, {Gand hi},
  {Pinhas}, {Madhusudhan}, \& {Deming}}]{Sheppard2017}
{Sheppard}, K.~B., {Mandell}, A.~M., {Tamburo}, P., {et~al.} 2017, \apjl, 850,
  L32

\bibitem[{{Sing} {et~al.}(2019){Sing}, {Lavvas}, {Ballester}, {Lecavelier des
  Etangs}, {Marley}, {Nikolov}, {Ben-Jaffel}, {Bourrier}, {Buchhave}, {Deming},
  {Ehrenreich}, {Mikal-Evans}, {Kataria}, {Lewis}, {L{\'o}pez-Morales},
  {Garc{\'\i}a Mu{\~n}oz}, {Henry}, {Sanz-Forcada}, {Spake}, {Wakeford}, \&
  {PanCET Collaboration}}]{Sing2019}
{Sing}, D.~K., {Lavvas}, P., {Ballester}, G.~E., {et~al.} 2019, \aj, 158, 91

\bibitem[{{Snellen} {et~al.}(2010){Snellen}, {de Kok}, {de Mooij}, \&
  {Albrecht}}]{Snellen2010}
{Snellen}, I.~A.~G., {de Kok}, R.~J., {de Mooij}, E.~J.~W., \& {Albrecht}, S.
  2010, \nat, 465, 1049

\bibitem[{{Stangret} {et~al.}(2020){Stangret}, {Casasayas-Barris}, {Pall{\'e}},
  {Yan}, {S{\'a}nchez-L{\'o}pez}, \& {L{\'o}pez-Puertas}}]{Stangret2020}
{Stangret}, M., {Casasayas-Barris}, N., {Pall{\'e}}, E., {et~al.} 2020, \aap,
  638, A26

\bibitem[{{Tabernero} {et~al.}(2021){Tabernero}, {Zapatero Osorio}, {Allart},
  {Borsa}, {Casasayas-Barris}, {Demangeon}, {Ehrenreich}, {Lillo-Box}, {Lovis},
  {Pall{\'e}}, {Sousa}, {Rebolo}, {Santos}, {Pepe}, {Cristiani}, {Adibekyan},
  {Allende Prieto}, {Alibert}, {Barros}, {Bouchy}, {Bourrier}, {D'Odorico},
  {Dumusque}, {Faria}, {Figueira}, {G{\'e}nova Santos}, {Gonz{\'a}lez
  Hern{\'a}ndez}, {Hojjatpanah}, {Lo Curto}, {Lavie}, {Martins}, {Martins},
  {Mehner}, {Micela}, {Molaro}, {Nunes}, {Poretti}, {Seidel}, {Sozzetti},
  {Su{\'a}rez Mascare{\~n}o}, {Udry}, {Aliverti}, {Affolter}, {Alves}, {Amate},
  {Avila}, {Bandy}, {Benz}, {Bianco}, {Broeg}, {Cabral}, {Conconi}, {Coelho},
  {Cumani}, {Deiries}, {Dekker}, {Delabre}, {Fragoso}, {Genoni}, {Genolet},
  {Hughes}, {Knudstrup}, {Kerber}, {Landoni}, {Lizon}, {Maire}, {Manescau}, {Di
  Marcantonio}, {M{\'e}gevand}, {Monteiro}, {Monteiro}, {Moschetti}, {Mueller},
  {Modigliani}, {Oggioni}, {Oliveira}, {Pariani}, {Pasquini}, {Rasilla},
  {Redaelli}, {Riva}, {Santana-Tschudi}, {Santin}, {Santos}, {Segovia},
  {Sosnowska}, {Span{\`o}}, {Tenegi}, {Iwert}, {Zanutta}, \&
  {Zerbi}}]{Tabernero2021}
{Tabernero}, H.~M., {Zapatero Osorio}, M.~R., {Allart}, R., {et~al.} 2021,
  \aap, 646, A158

\bibitem[{{Talens} {et~al.}(2018){Talens}, {Justesen}, {Albrecht}, {McCormac},
  {Van Eylen}, {Otten}, {Murgas}, {Palle}, {Pollacco}, {Stuik}, {Spronck},
  {Lesage}, {Grundahl}, {Fredslund Andersen}, {Antoci}, \&
  {Snellen}}]{Talens2018}
{Talens}, G.~J.~J., {Justesen}, A.~B., {Albrecht}, S., {et~al.} 2018, \aap,
  612, A57

\bibitem[{{Tamuz} {et~al.}(2005){Tamuz}, {Mazeh}, \& {Zucker}}]{Tamuz2005}
{Tamuz}, O., {Mazeh}, T., \& {Zucker}, S. 2005, \mnras, 356, 1466

\bibitem[{{Tsiaras} {et~al.}(2018){Tsiaras}, {Waldmann}, {Zingales},
  {Rocchetto}, {Morello}, {Damiano}, {Karpouzas}, {Tinetti}, {McKemmish},
  {Tennyson}, \& {Yurchenko}}]{Tsiaras2018}
{Tsiaras}, A., {Waldmann}, I.~P., {Zingales}, T., {et~al.} 2018, \aj, 155, 156

\bibitem[{{von Essen} {et~al.}(2019){von Essen}, {Mallonn}, {Welbanks},
  {Madhusudhan}, {Pinhas}, {Bouy}, \& {Weis Hansen}}]{Essen2019}
{von Essen}, C., {Mallonn}, M., {Welbanks}, L., {et~al.} 2019, \aap, 622, A71

\bibitem[{{Watanabe} {et~al.}(2020){Watanabe}, {Narita}, \&
  {Johnson}}]{Watanabe2020}
{Watanabe}, N., {Narita}, N., \& {Johnson}, M.~C. 2020, \pasj, 72, 19

\bibitem[{{Yan} {et~al.}(2019){Yan}, {Casasayas-Barris}, {Molaverdikhani},
  {Alonso-Floriano}, {Reiners}, {Pall{\'e}}, {Henning}, {Molli{\`e}re}, {Chen},
  {Nortmann}, {Snellen}, {Ribas}, {Quirrenbach}, {Caballero}, {Amado},
  {Azzaro}, {Bauer}, {Cort{\'e}s Contreras}, {Czesla}, {Khalafinejad}, {Lara},
  {L{\'o}pez-Puertas}, {Montes}, {Nagel}, {Oshagh}, {S{\'a}nchez-L{\'o}pez},
  {Stangret}, \& {Zechmeister}}]{Yan2019}
{Yan}, F., {Casasayas-Barris}, N., {Molaverdikhani}, K., {et~al.} 2019, \aap,
  632, A69

\bibitem[{{Yan} \& {Henning}(2018)}]{Yan&Henning2018}
{Yan}, F. \& {Henning}, T. 2018, Nature Astronomy, 2, 714

\bibitem[{{Yan} {et~al.}(2020){Yan}, {Pall{\'e}}, {Reiners}, {Molaverdikhani},
  {Casasayas-Barris}, {Nortmann}, {Chen}, {Molli{\`e}re}, \&
  {Stangret}}]{Yan2020}
{Yan}, F., {Pall{\'e}}, E., {Reiners}, A., {et~al.} 2020, \aap, 640, L5

\bibitem[{{Yan} {et~al.}(2021{\natexlab{a}}){Yan}, {Reiners}, {Pall{\'e}},
  {Shulyak}, {Stangret}, {Molaverdikhani}, {Nortmann}, {Molli{\`e}re},
  {Henning}, {Casasayas-Barris}, {Cont}, {Chen}, {Czesla},
  {S{\'a}nchez-L{\'o}pez}, {L{\'o}pez-Puertas}, {Ribas}, {Quirrenbach},
  {Caballero}, {Amado}, {Galad{\`i}-Enr{\`i}quez}, {Khalafinejad}, {Lara},
  {Montes}, {Morello}, {Nagel}, {Sedaghati}, {Zapatero Osorio}, \&
  {Zechmeister}}]{Yan_2021_submitted}
{Yan}, F., {Reiners}, A., {Pall{\'e}}, E., {et~al.} 2021{\natexlab{a}}, \aap,
  submitted

\bibitem[{{Yan} {et~al.}(2021{\natexlab{b}}){Yan}, {Wyttenbach},
  {Casasayas-Barris}, {Reiners}, {Pall{\'e}}, {Henning}, {Molli{\`e}re},
  {Czesla}, {Nortmann}, {Molaverdikhani}, {Chen}, {Snellen}, {Zechmeister},
  {Huang}, {Ribas}, {Quirrenbach}, {Caballero}, {Amado}, {Cont},
  {Khalafinejad}, {Khaimova}, {L{\'o}pez-Puertas}, {Montes}, {Nagel}, {Oshagh},
  {Pedraz}, \& {Stangret}}]{Yan2021}
{Yan}, F., {Wyttenbach}, A., {Casasayas-Barris}, N., {et~al.}
  2021{\natexlab{b}}, \aap, 645, A22

\bibitem[{{Zechmeister} {et~al.}(2014){Zechmeister}, {Anglada-Escud{\'e}}, \&
  {Reiners}}]{Zechmeister2014}
{Zechmeister}, M., {Anglada-Escud{\'e}}, G., \& {Reiners}, A. 2014, \aap, 561,
  A59

\end{thebibliography}

\appendix

\section{Parameters of planetary systems}
\label{Parameters of planetary systems}

\begin{table}[h]
        \caption{Parameters of the \object{WASP-33} and \object{KELT-20/MASCARA-2} systems used in this work.}             
        \label{tab-parameters}                           
        \centering                                       
        \renewcommand{\arraystretch}{1.15} 
        \begin{threeparttable}
                \begin{tabular}{l   l   l}                       
                        \noalign{\smallskip}
                        \hline\hline                             
                        \noalign{\smallskip}
                        Parameter (Unit) & WASP-33b & KELT-20b \\     
                        \noalign{\smallskip}
                        \hline                                   
                        \noalign{\smallskip}
                        \textit{Planet} & & \\ 
                        \noalign{\smallskip}
                        $R_\mathrm{p}$ ($R_\mathrm{Jupiter}$)     & $1.679_{-0.030}^{+0.019}$ \tablefootmark{a} & $1.83\pm0.07$ \tablefootmark{g} \\
                        $P_\mathrm{orb}$ (d)                & 1.219870897 \tablefootmark{b}               & 3.4741070 \tablefootmark{h}   \\
                        $T_\mathrm{0}$ (d)                  & 2454163.22449 \tablefootmark{b}             & 2457503.120049 \tablefootmark{h} \\
                        $\varv_\mathrm{sys}$ (km\,s$^{-1}$) & $-3.02\pm0.42$ \tablefootmark{c}            & $-24.48\pm0.04$ \tablefootmark{i} \\
                        $K_\mathrm{p}$ (km\,s$^{-1}$)       & $231\pm3$ \tablefootmark{a}                 & $173.4_{-1.5}^{+1.8}$ \tablefootmark{g} \\
                                                     &                                             & $169.3_{-4.6}^{+5.9}$ \tablefootmark{h} \\
                        $T_\mathrm{ingress}$ (d) \footnotemark{}             & $0.0124\pm0.0002$ \tablefootmark{d}         & $0.01996_{-0.00077}^{+0.00080}$ \tablefootmark{h} \\
                        $T_\mathrm{transit}$ (d) \footnotemark[\value{footnote}]             & $0.1143\pm0.0002$ \tablefootmark{d}         & $0.14898_{-0.00088}^{+0.00091}$ \tablefootmark{h} \\
                        log\,$g$ (cgs)                       & 3.46 \tablefootmark{d}                      & $<3.42$ \tablefootmark{h} \\
                        \noalign{\smallskip} \hline \noalign{\smallskip}
                        \textit{Star} & &  \\  
                        \noalign{\smallskip}
                        $R_*$ ($R_\mathrm{\sun}$)                             & $1.509_{-0.030}^{+0.016}$ \tablefootmark{a} & $1.60\pm0.06$ \tablefootmark{g}\\ 
                        $T_\mathrm{eff}$ (K)                                  & $7430\pm100$ \tablefootmark{e}              & $8980_{-130}^{+90}$ \tablefootmark{g}\\                     
                        $\varv_\mathrm{rot}\sin i_*$ (km\,s$^{-1}$)           &  $86.63_{-0.32}^{+0.37}$ \tablefootmark{f}  & $114\pm3$\tablefootmark{g} \\ 
                        \noalign{\smallskip}
                        \hline                                   
                \end{tabular}
                \tablefoot{
                        \tablefoottext{a}{\cite{Lehmann2015} with parameters from \cite{Kovacs2013}}, 
                        \tablefoottext{b}{\cite{Maciejewski2018}},
                        \tablefoottext{c}{\cite{Nugroho2017}},
                        \tablefoottext{d}{\cite{Kovacs2013}},
                        \tablefoottext{e}{\cite{Collier-Cameron2010}},
                        \tablefoottext{f}{\cite{Johnson2015}},
                        \tablefoottext{g}{\cite{Talens2018}},           
                        \tablefoottext{h}{\cite{Lund2017} -- we assumed a value of log\,$g$\,=\,3.0 to compute the model spectra because only an upper limit is reported},
                        \tablefoottext{i}{\cite{Rainer2021}.

                        }
                        
                }
        \end{threeparttable}      
\end{table}

\footnotetext{WASP-33b is subject to a rapid orbital precession. The transit chord, transit duration, and ingress duration therefore change with time \citep{Johnson2015, Watanabe2020, Cauley2021}.}

\section{Removal of telluric and stellar lines with \texttt{SYSREM}}
\label{Removal of telluric and stellar lines with SYSREM}

\texttt{SYSREM} is a detrending algorithm that was originally designed to remove systematic effects from transit light curves \citep{Tamuz2005}. In its application to the search for exoplanet atmospheres, each wavelength bin of the spectral matrix is treated as an individual light curve. The algorithm models the systematics by iteratively fitting the trend of each wavelength bin as a function of time. Subsequently, the model is subtracted from the data. Systematic effects can have multiple causes, such as variations in airmass, atmospheric water vapor along the line of sight, seeing, or instrumental effects.

We implemented \texttt{SYSREM} following the method described by \cite{Gibson2020}, which runs the algorithm in flux space instead of magnitude space \citep{Tamuz2005}. In a first step, we ran the algorithm in the traditional way, which resulted in a model-subtracted residual matrix for each \texttt{SYSREM} iteration. Then we divided the original spectral matrix by the sum of the models from each \texttt{SYSREM} iteration. We also divided the uncertainties by the final model for error propagation. 

The \texttt{SYSREM} implementation proposed by \cite{Gibson2020} comes with the advantage of preserving the strength of the planetary spectral signature falling onto stellar and telluric lines. The mathematical description of the normalized spectra is $1 + F_p/F_*$, where $F_p/F_*$ is the planet-to-star flux ratio.

\begin{figure}
        \centering
        \includegraphics[width=0.5\textwidth]{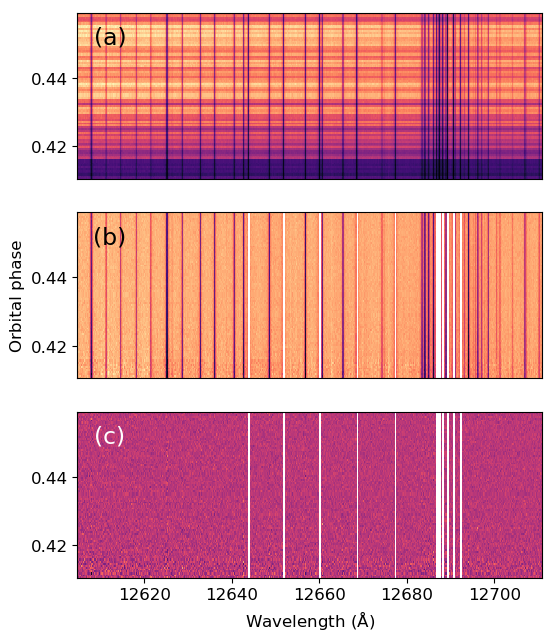}
        \caption{Example of data reduction steps for a selected CARMENES wavelength range (observation on 21 May 2020). {\it Panel~a} shows the unprocessed one-dimensional spectra. {\it Panel~b} illustrates the spectra after normalization and outlier correction; the strongest telluric lines are masked in this step. {\it Panel~c} shows the \texttt{SYSREM} reduced spectra after telluric and stellar line removal.}
        \label{SYSREM}
\end{figure}

%

\section{Additional figures}

\begin{figure}[h]
        \centering
        \includegraphics[width=0.5\textwidth]{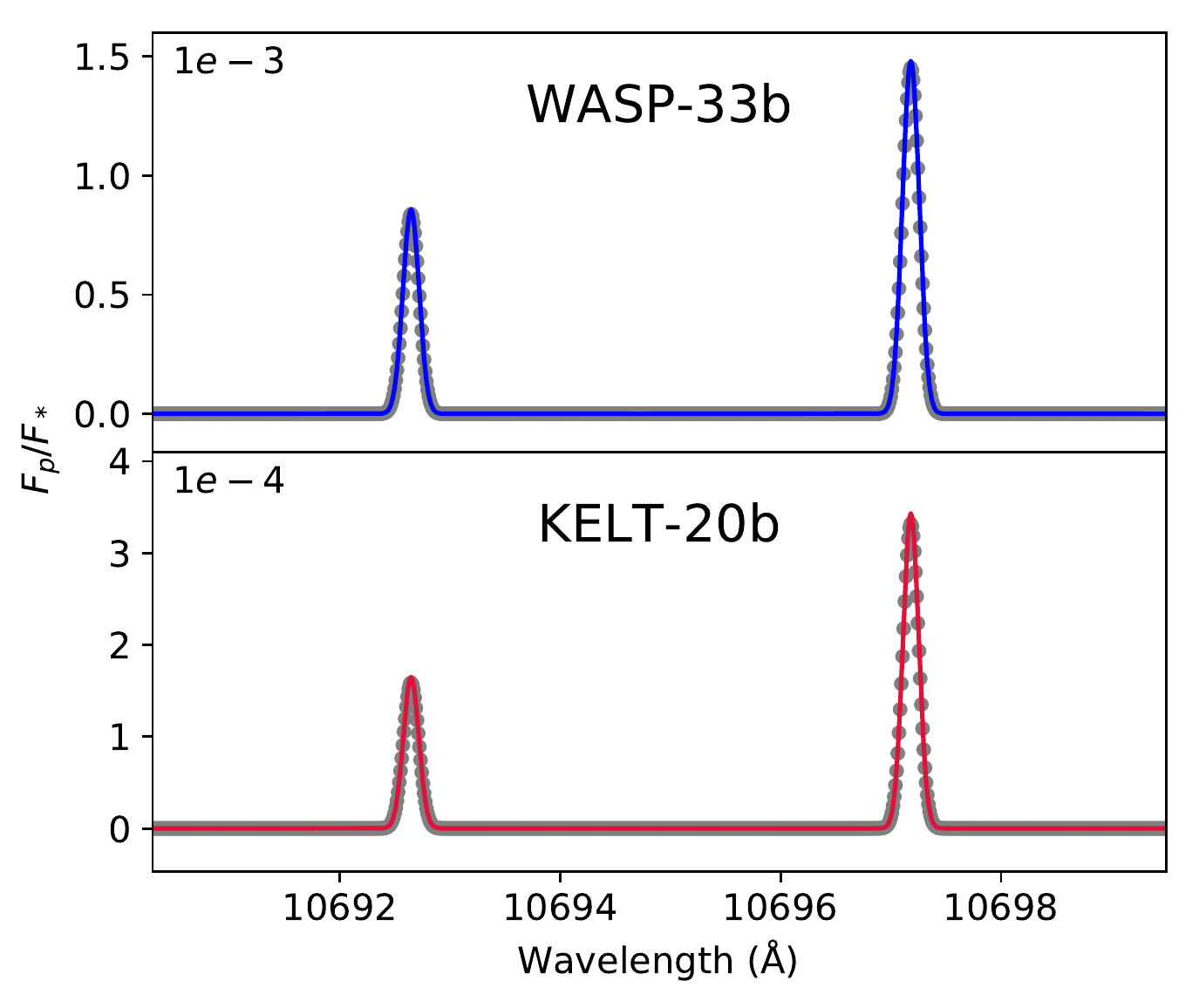}
        \caption{Comparison between model spectra with (gray data points) and without (solid lines) H$^-$ opacity. The difference between the models is insignificant, and hence the H$^-$ continuum opacity can be neglected.}
        \label{model_spectra_compare_H}
\end{figure}

\begin{figure*}
        \centering
        \includegraphics[width=\textwidth]{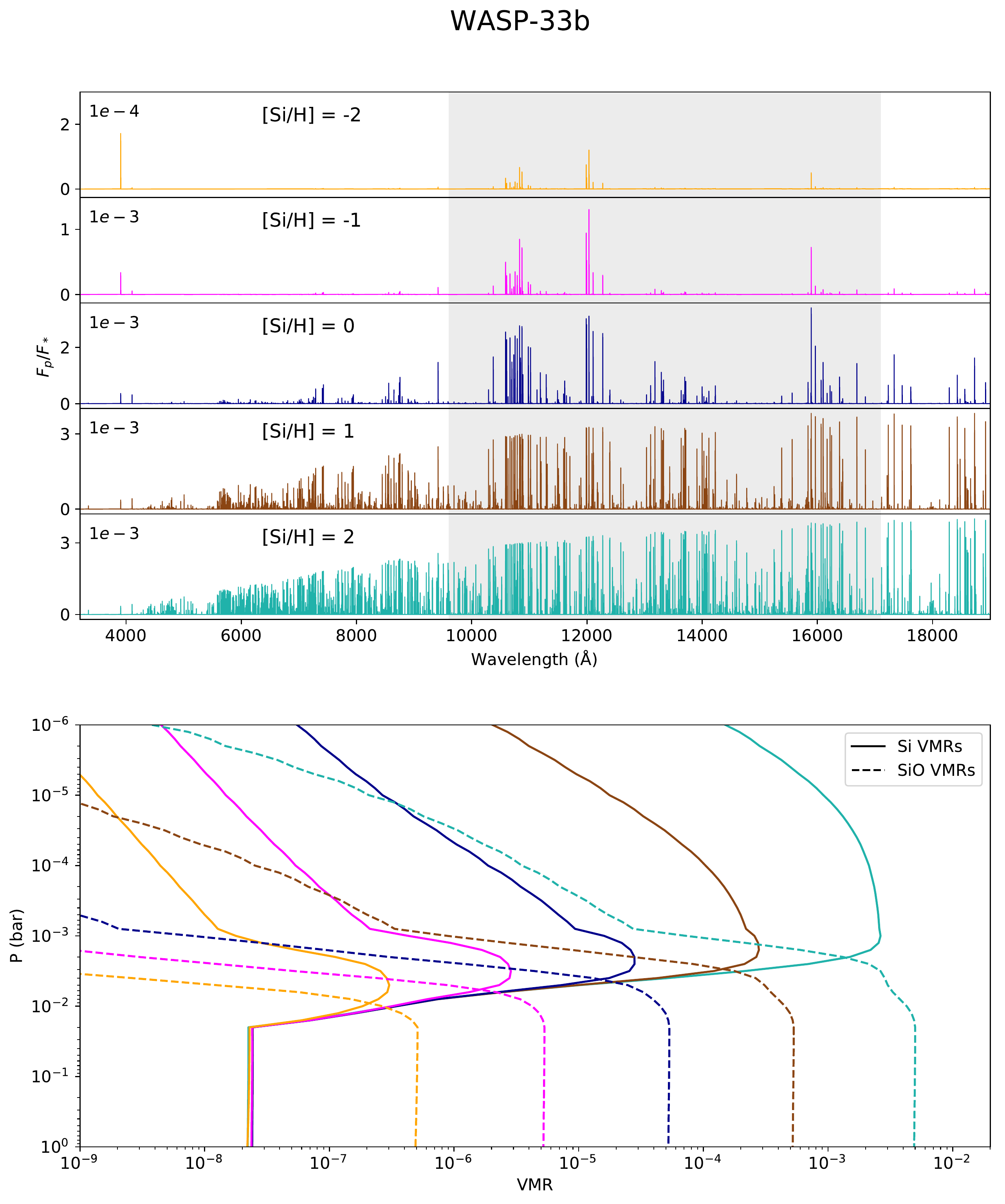}    
        \caption{Model spectra and VMRs at different metallicity values. {\it Top panel}: Model spectra for WASP-33b over a wide wavelength range (3000--19,000\,$\AA$). The gray shaded area corresponds to the CARMENES near-infrared channel. The spectra were calculated for VMRs with [Si/H] between --2\,dex and +2\,dex in steps of 1\,dex. The model with [Si/H]\,=\,0 is also shown in Fig.~\ref{model_spectra}.  {\it Bottom panel:}  VMRs computed by assuming chemical equilibrium. We also plot the VMRs of SiO (dashed lines) to allow for a comparison with the VMRs of Si (solid lines).}
        \label{model_spectra_W33_appendix}
\end{figure*}

\begin{figure*}
        \centering
        \includegraphics[width=\textwidth]{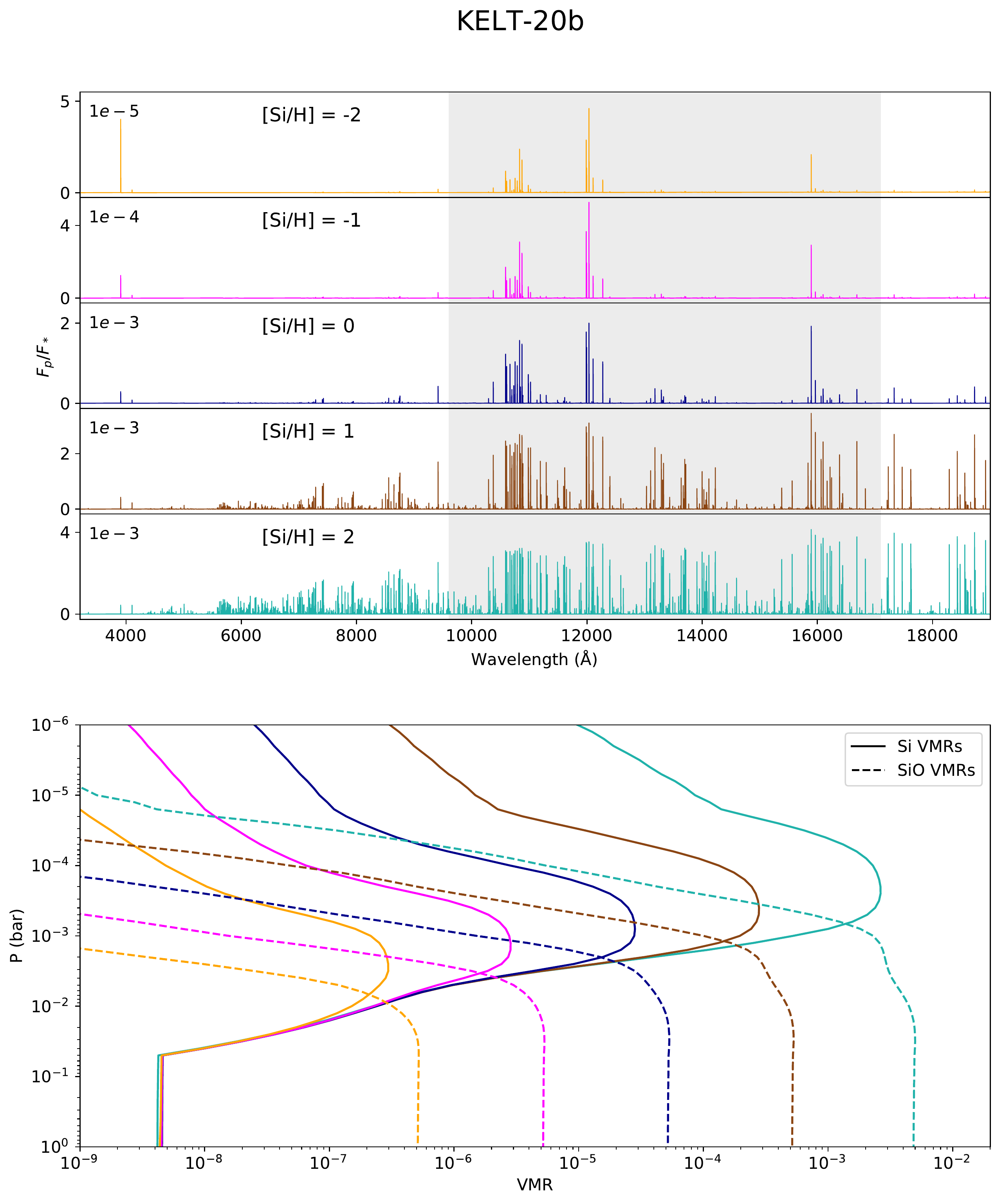}    
        \caption{Same as Fig.~\ref{model_spectra_W33_appendix}, but for KELT-20b.}
        \label{model_spectra_K20_appendix}
\end{figure*}

\begin{figure*}
        \centering
        \includegraphics[width=0.66\textwidth]{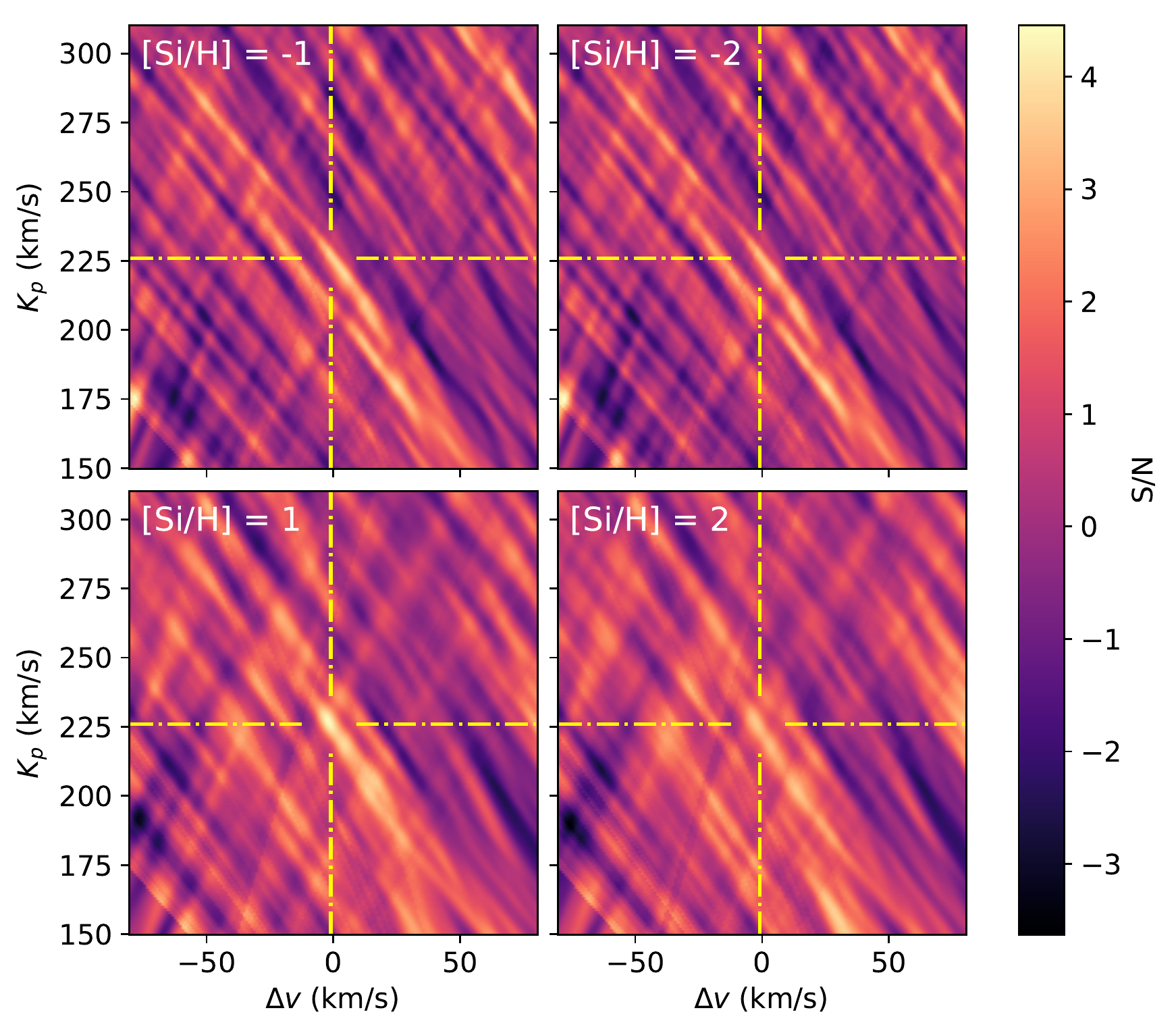}
        \caption{S/N maps of WASP-33b after three \texttt{SYSREM} iterations from model spectra with non-solar Si abundances. The {\it top panels} and {\it bottom panels} correspond to sub-solar and super-solar [Si/H] ratios, respectively. The yellow dashed-dotted lines indicate the location of the most significant detection peak, described in Sect.~\ref{Results and discussion}. The detection significance obtained with non-solar [Si/H] ratios is below the peak value obtained under the assumption of solar elemental abundances.}
        \label{SN_different_SiH_W33_appendix}
\end{figure*}

\begin{figure*}
        \centering
        \includegraphics[width=0.66\textwidth]{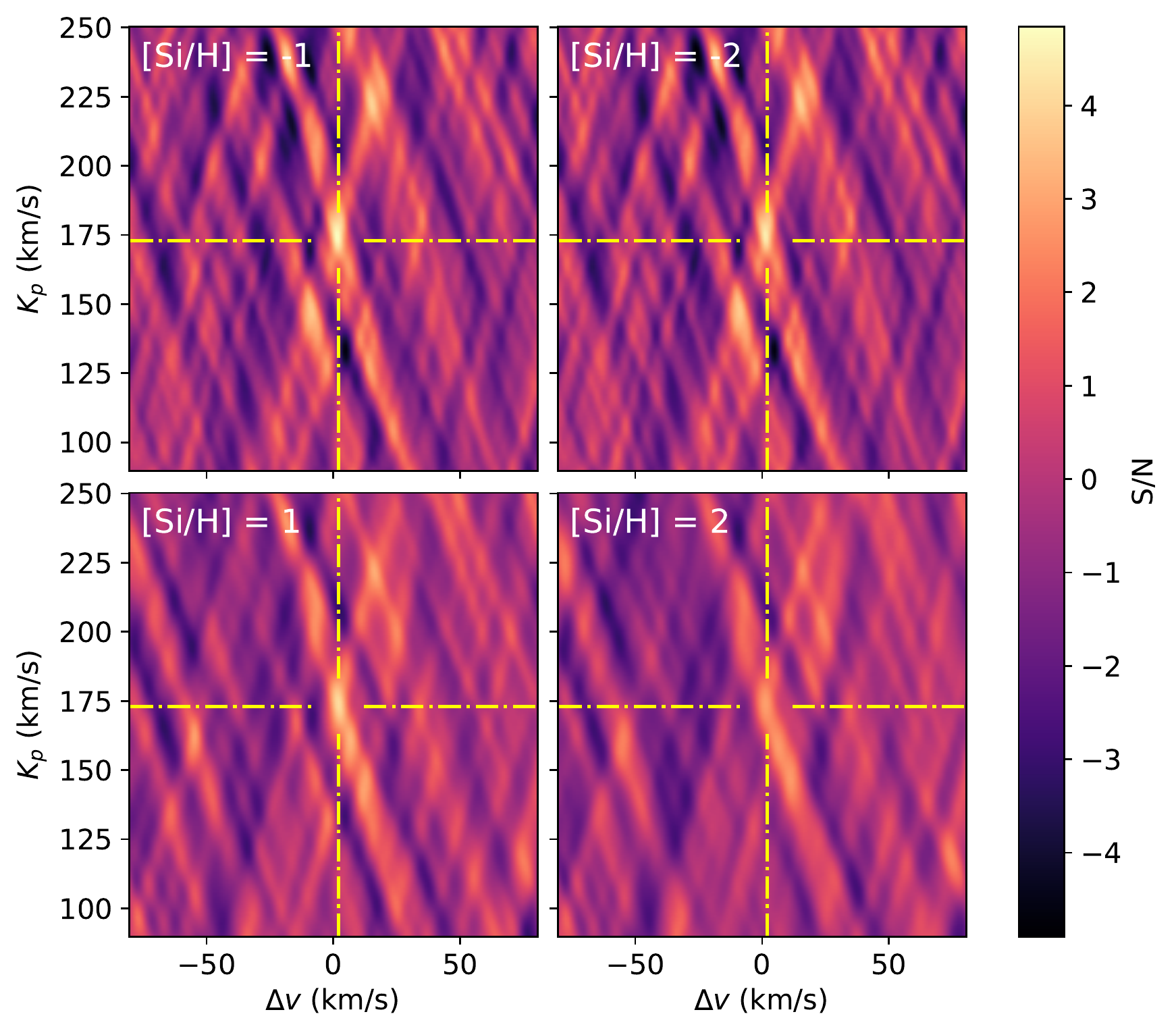}
        \caption{Same as Fig.~\ref{SN_different_SiH_W33_appendix}, but after four \texttt{SYSREM} iterations for KELT-20b.}
        \label{SN_different_SiH_K20_appendix}
\end{figure*}

\begin{figure*}
        \centering
        \includegraphics[width=\textwidth]{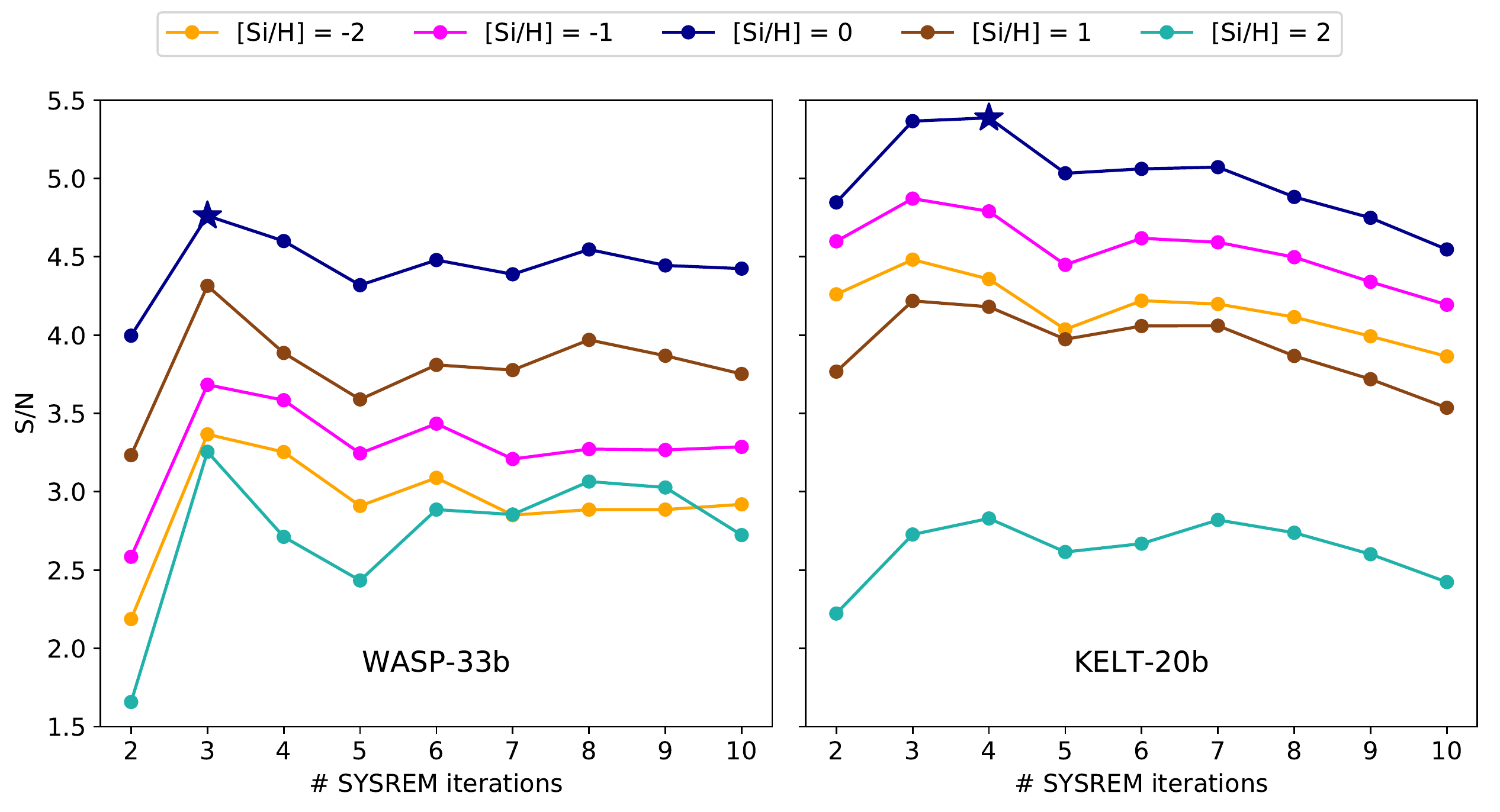}
        \caption{Evolution of the S/N detection strength (measured at the position of the strongest peak; see Sect.~\ref{Results and discussion}) with increasing \texttt{SYSREM} iteration. We show the S/Ns of WASP-33b and KELT-20b in the {\it left} and {\it right panels}, respectively. The different Si abundances are indicated by different colors. The strongest S/N peaks are found for both planets at [Si/H]\,=\,0 and are indicated by the blue stars.}
        \label{SN_iterations}
\end{figure*}

\begin{figure*}
        \centering
        \includegraphics[width=\textwidth]{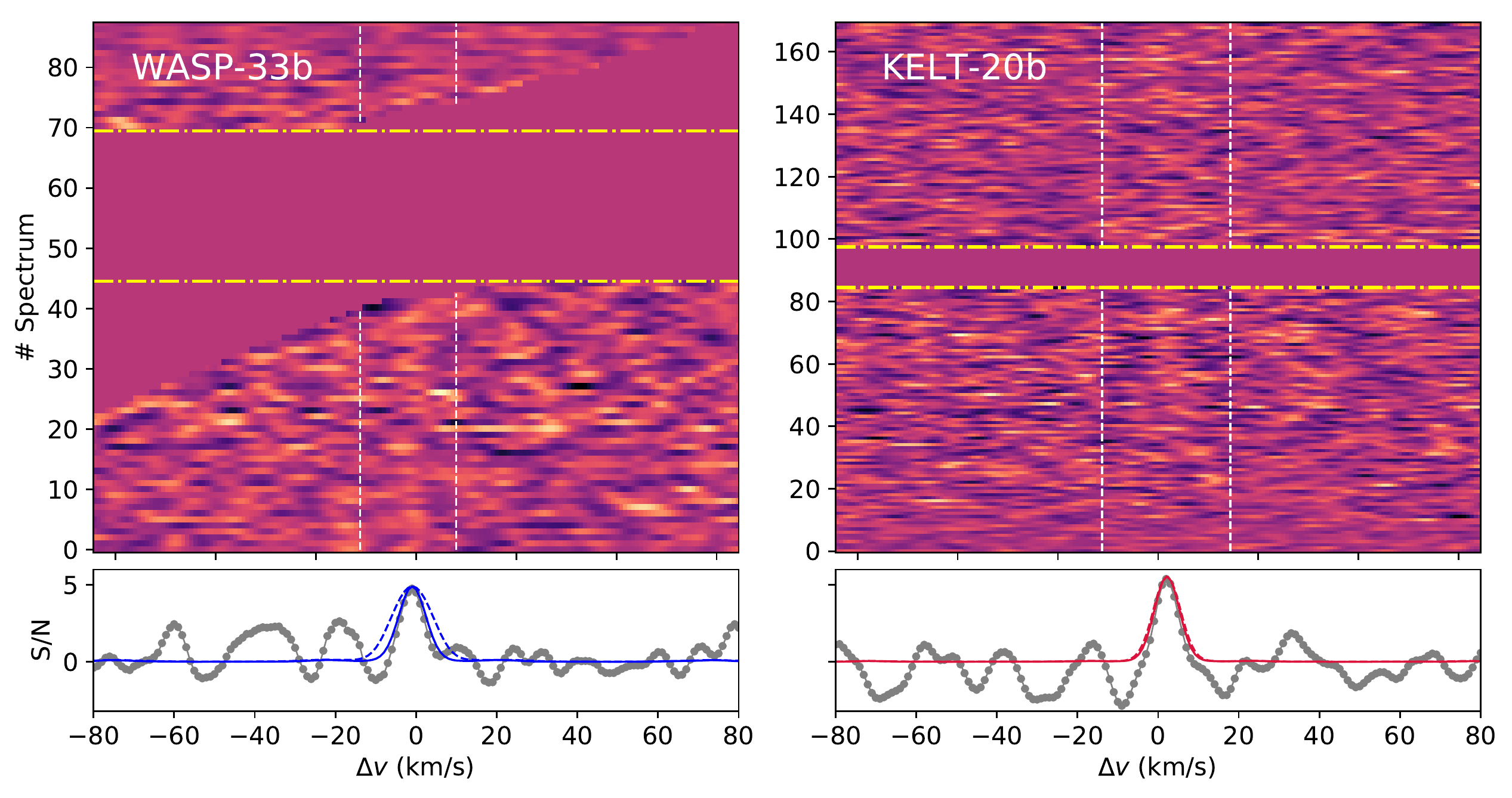} 
        \caption{CCF maps of WASP-33b ({\it left panels}) and KELT-20b ({\it right panels}). The aligned CCF maps are shown in the {\it top panels} (assuming $K_\mathrm{p}$ values of 226.0\,km\,s$^{-1}$ and 173.0\,km\,s$^{-1}$ for WASP-33b and KELT-20b, respectively). The vertical dashed lines indicate the planetary trail; the horizontal dashed-dotted lines indicate ingress and egress from secondary eclipse. As described in Sect.~\ref{Cross-correlation}, we masked the RV range of residual stellar Si lines in the CCF map of WASP-33b.
        In the {\it bottom panels} the collapsed CCF maps (gray lines) are compared to simulated CCFs (blue and red lines). The simulated CCFs that are rotationally broadened (by 7\,km\,s$^{-1}$ and 3\,km\,s$^{-1}$, respectively) are represented by the dashed lines. Those without broadening are represented by the solid lines. We note that in the case of KELT-20b, the simulated CCFs with and without broadening differ only marginally and therefore lie on top of each other.}
        \label{CCF_trails}
\end{figure*}

%

\clearpage

\section{Null detection test}
\label{Null detection test}

\begin{figure}[H]
        \centering
        \includegraphics[width=0.5\textwidth]{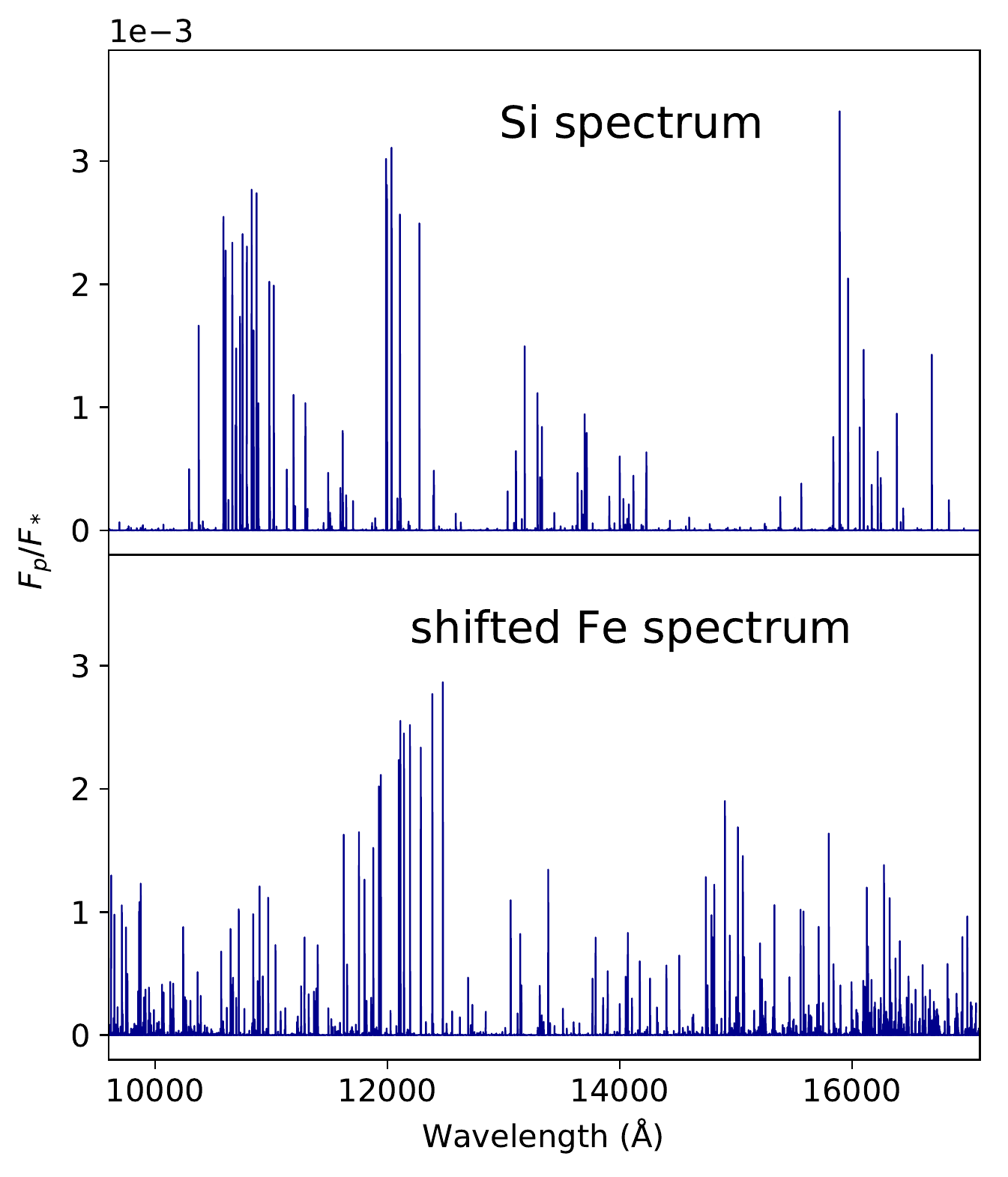}    
        \caption{Comparison between the Si model spectrum of WASP-33b ({\it top panel}) and the wavelength-shifted Fe model ({\it bottom panel}). We note that the density and amplitude of the emission lines in the two models are similar.}
        \label{null_test_spectra}
\end{figure}

To further increase confidence in the detected Si signals, we tested how the use of an inappropriate model spectrum for cross-correlation affects the S/N maps. We chose to use a shifted Fe model spectrum, since the amplitude and density of the Fe emission lines are similar to those of Si in the near-infrared wavelength range of CARMENES (see Fig.~\ref{null_test_spectra}). To avoid detecting a Fe signal (Fe is present in the atmosphere of WASP-33b and KELT-20b; \citealt{Nugroho2020_Fe}, \citealt{Cont2021}, \citealt{Yan2021}), the wavelength solution of the Fe model spectrum was shifted by a constant value of 500\,\AA. By shifting the wavelength axis, the Fe lines are no longer located at the right position. Hence, we reached a situation that corresponds to that of a model with random lines. We also tested shifts other than 500\,\AA, all of which led to the same conclusions.

\begin{figure}
        \centering
        \includegraphics[width=0.5\textwidth]{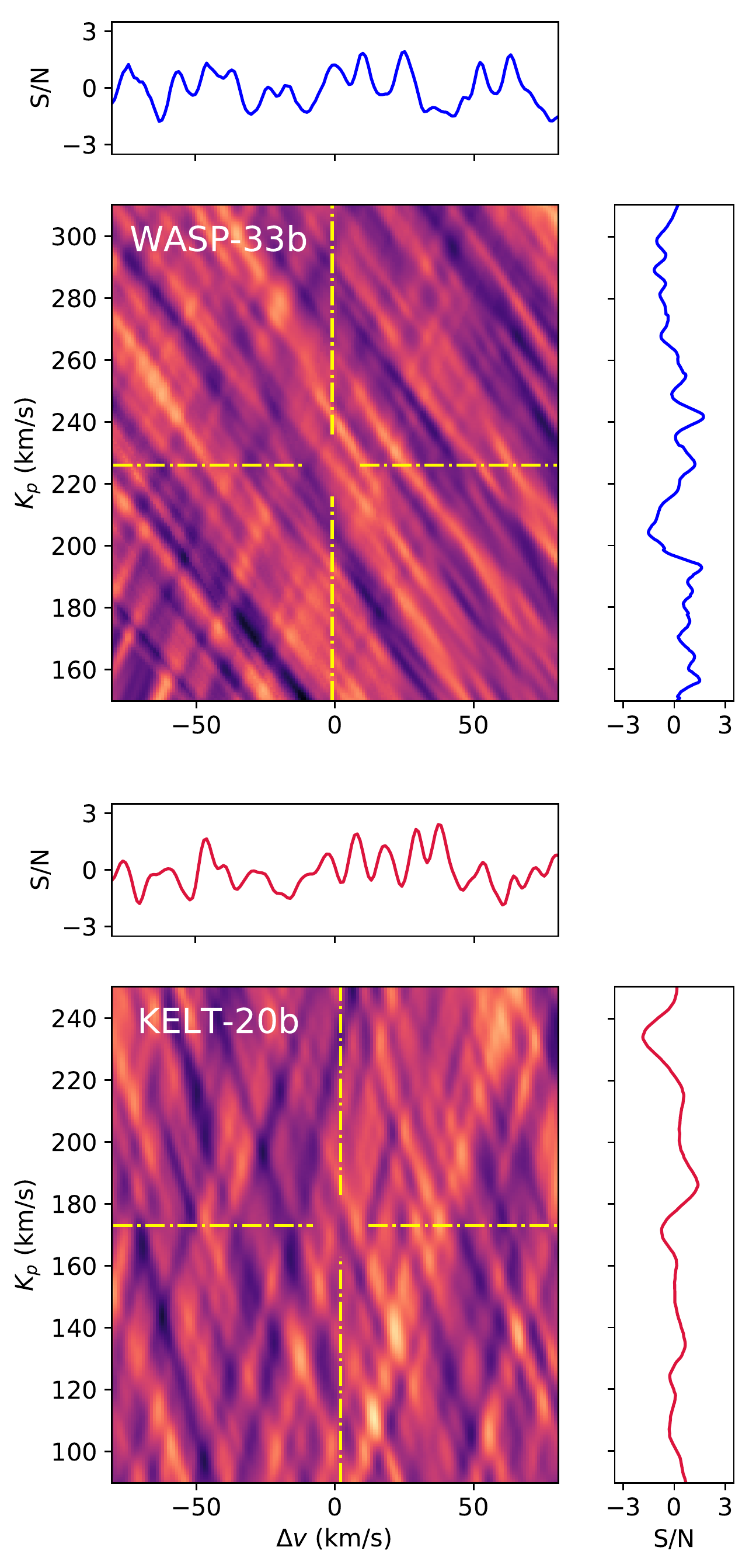}    
        \caption{S/N detection maps of WASP-33b ({\it top panel}) and KELT-20b ({\it bottom panel}), obtained from cross-correlation with the shifted Fe model spectrum. A random noise pattern without any significant detection peaks is found. The detection coordinates of Si are indicated by the yellow dashed-dotted lines. The horizontal and vertical panels correspond to the cross sections at the location of the Si S/N peaks and do not show any detection signal.}
        \label{null_test_SN_maps}
\end{figure}

We computed the CCFs and the S/N detection maps by using the shifted Fe model spectrum. The resulting S/N maps show a noise pattern without any strong detection peaks (see Fig.~\ref{null_test_SN_maps}). We conclude that our implementation of the cross-correlation technique does not lead to significant detections when an inappropriate model spectrum is used.

\newpage

\end{document}